\documentclass{CSML}

\def\dOi{12(1:5)2016}
\lmcsheading%
{\dOi}
{1--30}
{}
{}
{Apr.~29, 2014}
{May.~\phantom03, 2016}
{}


\ACMCCS{[{\bf Theory of computation}]: Logic---Proof theory; Computational complexity and cryptography---Proof complexity.}
\usepackage[utf8]{inputenc}
\usepackage[pdftex,dvipsnames,usenames]{color}
\usepackage[lutzsyntax,noxy]{virginialake}
\usepackage{upgreek}
\usepackage{hyperref}

\renewcommand{\le}{\leqslant}
\renewcommand{\ge}{\geqslant}

\begin{document}

\title[Quasipolynomial Normalisation in Deep Inference]
      {Quasipolynomial Normalisation in Deep Inference\\ 
       via Atomic Flows and Threshold Formulae}

\author[P.~Bruscoli]{Paola Bruscoli\rsuper a}
\address{{\lsuper{a,b}}University of Bath}
\thanks{{\lsuper{a,b}}EPSRC grants EP/E042805/1 \emph{Complexity and Non-determinism in Deep Inference} and EP/K018868/1 \emph{Efficient and Natural Proof Systems}; \emph{Senior Chaire d'Excellence `Identity and Geometric Essence of Proofs'} of the \emph{Agence Nationale de la Recherche}.}
\urladdr{http://www.cs.bath.ac.uk/pb/, http://alessio.guglielmi.name}
%\email{\{P.Buscoli, A.Guglielmi\}@Bath.Ac.UK}

\author[A.~Guglielmi]{Alessio Guglielmi\rsuper b}
\address{\vspace{-18 pt}}

\author[T.~Gundersen]{Tom Gundersen\rsuper c}
\address{{\lsuper c}Gundersen: Red Hat, Inc.}
\thanks{{\lsuper c}\emph{Overseas Research Studentship} and \emph{Research Studentship}, from the University of Bath; \emph{Senior Chaire d'Excellence `Identity and Geometric Essence of Proofs'} and \emph{Project STRUCTURAL---`Structural and Computational Proof Theory'} of the \emph{Agence Nationale de la Recherche}.}
\email{teg@jklm.no}

\author[M.~Parigot]{Michel Parigot\rsuper d}
\address{{\lsuper d}Laboratoire PPS, UMR 7126, CNRS \& Université Paris 7 (France).}
\thanks{{\lsuper d}\emph{Project INFER---`Theory and Application of Deep Inference'} and  \emph{Project STRUCTURAL---`Structural and Computational Proof Theory'} of the \emph{Agence Nationale de la Recherche}.}
\email{michel.parigot@gmail.com}

\begin{abstract}
\noindent Jeřábek showed that cuts in classical propositional logic proofs in deep inference can be eliminated in quasipolynomial time. The proof is indirect and it relies on a result of Atserias, Galesi and Pudlák about monotone sequent calculus and a correspondence between that system and cut-free deep-inference proofs. In this paper we give a direct proof of Jeřábek's result: we give a quasipolynomial-time cut-elimination procedure for classical propositional logic in deep inference. The main new ingredient is the use of a computational trace of deep-inference proofs called atomic flows, which are both very simple (they only trace structural rules and forget logical rules) and strong enough to faithfully represent the cut-elimination procedure.
\end{abstract}

\maketitle

%===============================================================================
\section{Introduction}

Deep inference is a proof-theoretic methodology where proofs can be freely composed by the logical operators, instead of having a rigid formula-directed tree structure, as in Gentzen proof
theory
\cite{Gugl:06:A-System:kl,
      BrunTiu:01:A-Local-:mz,
      Brun:04:Deep-Inf:rq,
      GuglGundPari::A-Proof-:fk}. As a result, inference rules apply arbitrarily deep inside formulae, contrary to traditional proof systems such as natural deduction and the sequent calculus, where inference rules only deal with the outermost structure of formulae. This induces a new symmetry, which can be exploited for achieving locality of inference rules, and which is not available with Gentzen methods. Locality, in turn, makes it possible to use new methods, often with a geometric flavour, in the normalisation theory of proof systems.

The greater freedom in composing proofs of deep inference is both a source of immediate technical difficulty and of new powerful proof-theoretic methods. A general methodology allows us to design deep-inference proof systems having more symmetries and finer structural properties than the sequent calculus does. For instance, cut and identity become really dual of each other, whereas they only are morally so in the sequent calculus, and all the structural rules can be reduced to their atomic form, whereas this does not hold in the sequent calculus, for example in the case of the contraction inference rule. In deep inference, the cut rule is more general than its counterpart in the sequent calculus, and makes it possible to obtain a broader range of dynamics in normalisation procedures. However, despite the sequent calculus systems and their normalisation procedures being special cases of deep inference systems and procedures, cut elimination in deep inference still guarantees consistency and the trivial turning of proof systems into algorithms for proof search.

\newcommand{\SKS}{\mathsf{SKS}}
Sixteen years of research have guaranteed that all usual logics have deep-inference proof systems enjoying cut elimination (see
\cite{Gugl::Deep-Inf:uq} for a complete overview). The traditional methods of cut elimination of the sequent calculus can be adapted to a large extent to deep inference, despite having to cope with a higher generality, but new methods are also achievable. The standard proof system for propositional classical logic in deep inference is system $\SKS$, and its cut elimination has been achieved in several different ways
\cite{BrunTiu:01:A-Local-:mz,
      Brun:04:Deep-Inf:rq,
      GuglGund:07:Normalis:lr}, all requiring at worst exponential time in the size of the proof to be normalised.

A few years ago Jeřábek showed that cut elimination in $\SKS$ proofs can be done in quasipolynomial time
\cite{Jera::On-the-C:kx}, more specifically in time $n^{O(\log n)}$. The result is surprising because all known cut-elimination methods for classical-logic proof systems require exponential time, in particular for Gentzen's sequent calculus. Jeřábek obtained his result by relying on a construction over threshold functions by Atserias, Galesi and Pudlák, in the monotone sequent calculus
\cite{AtseGalePudl:02:Monotone:yu}.

Jeřábek's technique is indirect because normalisation is performed over proofs in the sequent calculus, which are, in turn, related to deep-inference ones by polynomial simulations, originally studied in
\cite{Brun:06:Deep-Inf:qy} and
\cite{BrusGugl:07:On-the-P:fk}.

In this paper we give a direct proof of Jeřábek's result: that is, we give a quasipoly\-no\-mial-time cut-elimination procedure in propositional-logic deep inference, which, in addition to being internal, has a strong computational flavour. Our proof uses two ingredients:
\begin{enumerate}
%---------------------------------------
\item an adaptation of Atserias-Galesi-Pudlák technique to deep inference,  which simplifies the technicalities associated with the use of threshold functions; in particular, the formulae and derivations that we adopt are smaller and structurally simpler than those in
\cite{AtseGalePudl:02:Monotone:yu};
%---------------------------------------
\item a recently introduced graphic formalism, tracing atoms in deep-inference proofs, called `atomic flows' 
\cite{GuglGund:07:Normalis:lr}.
\end{enumerate}

\noindent Atomic flows, which can be considered specialised Buss flow graphs
\cite{Buss:91:The-Unde:uq}, play a major role in designing and controlling the cut elimination procedure presented in this paper. Atomic flows are very simple (they only trace structural rules and forget logical rules) but they are strong enough to faithfully represent cut elimination
\cite{GuglGund:07:Normalis:lr,
      GuglGundStra::Breaking:uq} and to relate several formalisms regarding their proof complexity
\cite{Das:12:Complexi:kx,
      Das:13:The-Pige:fk}. Atomic flows contribute to the overall clarification of this highly technical matter, by reducing our dependency on syntax. The techniques developed via atomic flows tolerate variations in the proof system specification. In fact, their geometric nature makes them largely independent of syntax, provided that certain linearity conditions are respected (and this is usually achievable in deep inference).

The paper is self-contained. Sections~\ref{SectDI} and \ref{SectAF} are devoted, respectively, to the necessary background on deep inference and atomic flows. Threshold functions and formulae are introduced in Section~\ref{SectThresh}.

We normalise proofs in two steps, each of which has a dedicated section in the paper:
\begin{enumerate}
%---------------------------------------
\item We transform any given proof into what we call its `simple form'. No use is made of threshold formulae and no significant proof complexity is introduced. This is presented in Section~\ref{SectSimpleForm}, mostly an exercise on deep inference and atomic flows.
%---------------------------------------
\item In Section~\ref{SectPreNorm}, we show the cut elimination step, starting from proofs in simple form. Here, threshold formulae play a major role.
\end{enumerate}

\noindent Normalisation can be taken one step further, by removing the instances of the only inference rule left that is not analytic in the deep-inference sense, \emph{viz.}\ coweakening. This is performed by a simple and standard deep-inference procedure in Section~\ref{SectNorm}.

Section~\ref{SectFinComm} concludes the paper with comments on future research directions.

Parts of this paper were presented at LPAR 16
\cite{BrusGuglGundPari:09:A-Quasip:fk} and some appear in
\cite{Gund:09:A-Genera:kx}. Recently, threshold functions have been used in
\cite{Das:13:The-Pige:fk} to build quasipolynomial size cut-free deep-inference proofs of the propositional pigeonhole principle that, crucially, do not use cocontraction, which is a form of dagness.

%===============================================================================
\section{Propositional Logic in Deep Inference}\label{SectDI}

Inside the deep-inference methodology we can define several formalisms, \emph{i.e.} general prescriptions on how to design proof systems, in the same sense as the sequent calculus and natural deduction are formalisms in Gentzen-style proof theory (where the structure of proofs is determined by the tree structure of the formulae they prove).

The first, and conceptually simpler, formalism that has been defined in deep inference is called the \emph{calculus of structures}, or \emph{CoS}
\cite{Gugl:06:A-System:kl}. Another deep-inference formalism has later been introduced  in
\cite{GuglGundPari::A-Proof-:fk}, called \emph{open deduction}. Open deduction is more general than CoS, in the sense that every CoS derivation is also an open-deduction derivation. On the other hand, every open-deduction derivation can be transformed into a CoS derivation by a straightforward transformation that essentially amounts to interleaving derivations. The cost of this transformation is at most quadratic in the size of the original open-deduction derivation; therefore, from the point of view of complexity, CoS and open deduction are equivalent.

CoS and open deduction are equivalent also from the point of view of proof theory, because the two formalisms are just two different notations for derivations of the same nature, and so every derivation transformation that can be performed in one formalism can also be performed in the other. In this paper we will adopt the open-deduction notation, especially because it is more efficient for the reader. However, given that most of the literature in deep inference adopts the CoS notation, which is more similar to the traditional Gentzen syntax, we will present both styles in this section.

The standard proof system of propositional logic in deep inference is called $\SKS$. The basic proof-complexity properties of $\SKS$, and so of propositional logic in deep inference, have been studied in
\cite{BrusGugl:07:On-the-P:fk} (which also could be used as an introduction to $\SKS$). Those properties are:
\begin{itemize}
%---------------------------------------
\item $\SKS$ is polynomially equivalent to Frege proof systems.
%---------------------------------------
\item $\SKS$ can be extended with Tseitin's extension and substitution, and the proof systems so obtained are polynomially equivalent to Frege proof systems augmented with extension and substitution.
%---------------------------------------
\item Cut-free $\SKS$ polynomially simulates cut-free Gentzen proof systems for propositional logic, but the converse does not hold: in fact, Statman's tautologies admit polynomial proofs in cut-free $\SKS$ but only exponential ones in cut-free Gentzen
\cite{Stat:78:Bounds-f:fj}.
\end{itemize}

\noindent We now quickly introduce all the necessary notions. An excellent and more relaxed introduction to $\SKS$ in CoS and its basic properties is
\cite{Brun:04:Deep-Inf:rq}.

\newcommand{\fff}{\mathsf f}
\newcommand{\ttt}{\mathsf t}
\emph{Formulae}, denoted by $A$, $B$, $C$ and $D$ are freely built from: \emph{units}, $\fff$ (false), $\ttt$ (true); \emph{atoms}, denoted by $a$, $b$, $c$, $d$ and $e$; \emph{disjunction} and \emph{conjunction}, ${\vlsbr[A.B]}$ and $\vlsbr(A.B)$. The different brackets have the only purpose of improving legibility; we usually omit external brackets of formulae, and sometimes we omit superfluous brackets under associativity. On the set of atoms a (non-identical) involution $\bar\cdot$ is defined and called \emph{negation}; $a$ and $\bar a$ are \emph{dual} atoms. We denote \emph{contexts}, \emph{i.e.}, formulae with a hole, by $K\vlhole$ and $H\vlhole$; for example, if $K\vlscn[a]$ is $\vls(b.[a.c])$, then $K\vlhole$ is $\vls(b.[\vlhole.c])$, $K\{b\}$ is $\vls(b.[b.c])$ and $K\vlscn(a.d)$ is $\vls(b.[(a.d).c])$.

Note that negation is only defined for atoms, and this is not a limitation because, thanks to De Morgan laws, negation can always be `pushed to' atoms. Also, note that there are no negative or positive atoms in an absolute sense; we can only say that if we arbitrarily consider $\bar a$ positive, then $a$ must be negative, for example.

For both CoS and open deduction an (\emph{inference}) \emph{rule} $\rho$ is an expression $\vlupsmash{\vlinf\rho{}BA}$, where the formulae $A$ and $B$ are called \emph{premiss} and \emph{conclusion}, respectively; an \emph{instance} of that rule is an expression $\vlinf\rho{}DC$, where $C$ and $D$ are instances of $A$ and $B$.

\newcommand{\ai  }{\mathsf{ai}}
\newcommand{\aw  }{\mathsf{aw}}
\newcommand{\ac  }{\mathsf{ac}}
\newcommand{\aid }{{\ai{\downarrow}}}
\newcommand{\awd }{{\aw{\downarrow}}}
\newcommand{\acd }{{\ac{\downarrow}}}
\newcommand{\aiu }{{\ai{\uparrow}}}
\newcommand{\awu }{{\aw{\uparrow}}}
\newcommand{\acu }{{\ac{\uparrow}}}
\newcommand{\swi }{\mathsf{s}}
\newcommand{\med }{\mathsf{m}}
\emph{System\/ $\SKS$} is a proof system, common to CoS and open deduction, defined by the following \emph{structural} inference rules:
\[
\begin{array}{@{}c@{}c@{}c@{}}
      \vlinf\aid{}{\vls[a.{\bar a}]}\ttt&
\qquad\vlinf\awd{}a\fff                 &
\qquad\vlinf\acd{}a{\vls[a.a]}          \\
\noalign{\smallskip}
      \emph{identity}                   &
\qquad\emph{weakening}                  &
\qquad\emph{contraction}                \\
\noalign{\bigskip}
      \vlinf\aiu{}\fff{\vls(a.{\bar a})}&
\qquad\vlinf\awu{}\ttt a                &
\qquad\vlinf\acu{}{\vls (a.a)}a         \\
\noalign{\smallskip}
      \emph{cut}&
\qquad\emph{coweakening}&
\qquad\emph{cocontraction}\\
\end{array}
\quad,
\]
and by the following two \emph{logical} inference rules:
\[
\begin{array}{@{}c@{}c@{}}
\vlinf\swi{}{\vls[(A.B).C]}{\vls(A.[B.C])}        &\qquad
\vlinf\med{}{\vls([A.C].[B.D])}{\vls[(A.B).(C.D)]}\\
\noalign{\smallskip}
\emph{switch}                                     &\qquad
\emph{medial}                                     \\
\end{array}
\quad.
\]
A \emph{cut-free} derivation is a derivation where $\aiu$ is not used, \emph{i.e.}, a derivation in $\SKS\setminus\{\aiu\}$. In addition to these rules, there is a rule $\vldownsmash{\vlinf={}DC}$, such that $C$ and $D$ are opposite sides in one of the following equations:
\begin{equation}\label{Eq}
\begin{array}{@{}r@{}l@{}r@{}l@{}}
\vls[A.B]      &{}=\vls[B.A]    &\qquad\qquad
\vls[A.\fff]   &{}=\vls[A]      \\
\noalign{\smallskip}
\vls(A.B)      &{}=\vls(B.A)    &\qquad\qquad
\vls(A.\ttt)   &{}=\vls(A)      \\
\noalign{\smallskip}
\vls[[A.B].C]  &{}=\vls[A.[B.C]]&\qquad\qquad
\vls[\ttt.\ttt]&{}=\vls[\ttt]   \\
\noalign{\smallskip}
\vls((A.B).C)  &{}=\vls(A.(B.C))&\qquad\qquad
\vls(\fff.\fff)&{}=\vls(\fff)                 
\end{array}
\quad.
\end{equation}
We do not always show the instances of rule $=$, and when we do show them, we gather several contiguous instances into one. We consider the $=$ rule as implicitly present in all systems. The equality relation $=$ on formulae is defined by closing the equations in \eqref{Eq} by reflexivity, symmetry, transitivity and by stipulating that $A=B$ implies $K\vlscn[A]=K\vlscn[B]$; to indicate literal equality of the formulae $A$ and $B$ we adopt the notation $A\equiv B$.

We now define both styles of derivations, CoS and open deduction. The difference is in the way we compose instances of rules: in CoS we only allow inferences to compose vertically, in chains similar to sequent calculus proofs made only of one-premiss rule instances. In open deduction instead, derivations can be composed by the same connectives that formulae are made of. For simplicity, we give here a definition of open-deduction derivation that is limited to our purposes in this paper, and is not the most general.

In CoS, a rule instance $\vlinf\rho{}DC$ generates an (\emph{inference}) \emph{step} $\vlinf\rho{}{K\vlscn[D]}{K\vlscn[C]}$, for each context $K\vlhole$. A CoS \emph{derivation} \emph{from (premiss) $A$ to (conclusion) $B$}  is a chain of inference steps with $A$ at the top and $B$ at the bottom. A derivation can be denoted by
\[
\vlder\Phi{\mathcal S}BA
\quad,
\]
where $\mathcal S$ is the name of the proof system or a set of inference rules (we might omit $\Phi$ and $\mathcal S$); a \emph{proof}, often denoted by $\Pi$, is a derivation with premiss $\ttt$; besides $\Phi$, we denote derivations with $\Psi$. Sometimes we group $n\ge0$ inference steps of the same rule $\rho$ together into one step, and we label the step with $n\cdot\rho$.

In open deduction, and just for the specific case of propositional logic with $\vlor$, $\vlan$ and negation on atoms, we define the notions of \emph{derivation}, \emph{premiss} and \emph{conclusion} inductively as follows:
\begin{itemize}
%---------------------------------------
%---------------------------------------
\item if $\Phi$ is a unit or an atom then $\Phi$ is a derivation with premiss $\Phi$ and conclusion $\Phi$;
%---------------------------------------
%---------------------------------------
\item if $\Phi$ is a derivation with premiss $A$ and conclusion $B$ and if $\Psi$ is a derivation with premiss $C$ and conclusion $D$, then
\begin{itemize}
%---------------------------------------
\item $\vlsbr[\Phi.\Psi]$ is a derivation with premiss $\vlsbr[A.C]$ and conclusion $\vlsbr[B.D]$,
%---------------------------------------
\item $\vlsbr(\Phi.\Psi)$ is a derivation with premiss $\vlsbr(A.C)$ and conclusion $\vlsbr(B.D)$,
%---------------------------------------
\item if $\vlinf\rho{}CB$ is an instance of an inference rule, then $\vlinf\rho{}\Psi\Phi$ is a derivation with premiss $A$ and conclusion $D$.
\end{itemize}
\end{itemize}
We adopt the same conventions as for CoS to denote derivations in open deduction. We omit structural rule names in open-deduction notation.

The first two rows in Figure~\ref{FigExAF} illustrate with examples all the concepts introduced above. The first row shows three example CoS derivations, and below each of them there is an equivalent derivation in open deduction. An open deduction derivation can be obtained from a CoS one by sharing the contexts in inference steps. Vice versa, a CoS derivation can be obtained from an open deduction one by choosing an order for the chain of inference steps.

\newcommand  {\SKSg}{\mathsf{SKSg}}
\newcommand  {\gw  }{\mathsf w}
\newcommand  {\gwd }{{\gw{\downarrow}}}
\newcommand  {\gwu }{{\gw{\uparrow}}}
\newcommand  {\gc  }{\mathsf c}
\renewcommand{\gcd }{{\gc{\downarrow}}}
\newcommand  {\gcu }{{\gc{\uparrow}}}
\newcommand  {\gi  }{\mathsf i}
\newcommand  {\gid }{{\gi{\downarrow}}}
Besides $\SKS$, another standard deep-inference system is $\SKSg$, which is the same as $\SKS$, except that it does not contain medial and its structural rules are not restricted to atoms. In particular, we use in this paper the rules
\[
\vlinf\gwd{}A\fff
\quad,\qquad
\vlinf\gwu{}\ttt A
\quad,\qquad
\vlinf\gcd{}A{\vls[A.A]}
\qquad\text{and}\qquad
\vlinf\gcu{}{\vls(A.A)}A
\quad.
\]
Clearly, a derivation in $\SKS$ is also a derivation in $\SKSg$. It can easily be proved that $\SKS$ and all its fragments containing the logical and $=$ rules polynomially simulate, respectively, $\SKSg$ and its corresponding fragments
\cite{BrusGugl:07:On-the-P:fk}. For example, $\{\swi,\med,=,\acd\}$ polynomially simulates $\{\swi,=,\gcd\}$. This allows us to transfer properties from $\SKS$ to $\SKSg$; in particular, the main result in this paper, \emph{i.e.}, that $\SKS$ proofs can be transformed into cut-free ones in quasipolynomial time, holds also for $\SKSg$ proofs. One reason to work with $\SKS$ instead of $\SKSg$, as we do in this paper, is that atomicity of rules allows us to use atomic flows more conveniently.

\newcommand{\KS}{\mathsf{KS}}
A notable cut-free system is $\KS=\{\swi,\med,=,\aid,\awd,\acd\}$, which is complete for propositional logic
\cite{BrunTiu:01:A-Local-:mz,
      Brun:04:Deep-Inf:rq}; this, of course, entails completeness for all the systems that contain $\KS$, such as $\SKS$.

We can replace instances of nonatomic structural rules by derivations with the same premiss and conclusion, and that only contain atomic structural rules. The price to pay is a quadratic growth in size. This is stated by the following, routine proposition (keep in mind that, from now on, we consider the $=$ rule as implicitly present in all systems). An example is the rightmost upper derivation in Figure~\ref{FigExAF}, which stands for a nonatomic cocontraction.

%-------------------------------------------------------------------------------
\begin{prop}\label{PropGenAtPol}
Rule instances of\/ $\gwd$, $\gwu$, $\gcd$ and\/ $\gcu$ can be derived in quadratic time by derivations in\/ $\{\awd\}$, $\{\awu\}$, $\{\med,\acd\}$ and\/ $\{\med,\acu\}$, respectively.
\end{prop}

Sometimes, we use a nonatomic rule instance to stand for some derivation in $\SKS$ that derives that instance, as per Proposition~\ref{PropGenAtPol}.

By $A\{a_1/B_1,\dots,a_h/B_h\}$, we denote the operation of simultaneously substituting formulae $B_1$, \dots, $B_h$ into all the occurrences of the atoms $a_1$, \dots, $a_h$ in the formula $A$, respectively; note that the occurrences of $\bar a_1$, \dots, $\bar a_h$ are not automatically substituted. Often, we only substitute certain occurrences of atoms, and these are indicated with superscripts that establish a relation with atomic flows. As a matter of fact, we extend the notion of substitution to derivations in the natural way, but this requires a certain care. The issue is clarified in Section~\ref{SectAF} (see, in particular, Notations~\ref{NotSubst} and \ref{NotDerSubst} and Proposition~\ref{PropSubst}).

\newcommand{\size}[1]{{\left\vert #1\right\vert}}\vlupdate\size
The \emph{size} $\size A$ of a formula $A$, and the \emph{size} $\size\Phi$ of a derivation $\Phi$, is the number of unit and atom occurrences appearing in it. 
The size of CoS derivations is, obviously, at most quadratic in the size of the corresponding open-deduction derivations. We use this fact implicitly throughout the paper, and we always measure the CoS size of derivations, even if we show them in open-deduction notation.

%===============================================================================
\section{Atomic Flows}\label{SectAF}

Atomic flows, which have been introduced in
\cite{GuglGund:07:Normalis:lr}, are, essentially, specialised Buss flow graphs
\cite{Buss:91:The-Unde:uq}. They are particular directed graphs associated with $\SKS$ derivations: every derivation yields one atomic flow obtained by tracing the atom occurrences in the derivation. Infinitely many derivations correspond to each atomic flow; this suggests that much of the information in a derivation is lost in its associated atomic flow; in particular, there is no information about instances of logical rules, only structural rules play a role. As shown in 
\cite{GuglGund:07:Normalis:lr,
      GuglGundStra::Breaking:uq}, it turns out that atomic flows contain sufficient structure to control normalisation procedures, providing in particular induction measures that can be used to ensure termination. Such normalisation procedures require exponential time on the size of the derivation to be normalised. In the present work, we lower the complexity of proof normalisation to quasipolynomial time, but an essential role is played by the complex logical relations of threshold formulae, which are external and independent from the given proof. This means that atomic flows are not sufficient to define the normalisation procedure; however, they still are a very convenient tool for defining and understanding several of its aspects.

We can single out three features of atomic flows that, in general, and not just in this work, help in designing normalisation procedures:

\begin{enumerate}
%---------------------------------------
\item\label{ItemUseTop} Atomic flows conveniently express the topological structure of atom occurrences in a proof. This is especially useful for defining a certain `simple form' of proofs (Definition~\ref{DefSimpleForm}).
%---------------------------------------
\item\label{ItemUseSubst} Atomic flows provide for an efficient way to control substitutions for atom occurrences in derivations. This helps us to define the cut-free form of proofs (Definition~\ref{DefNorm}).
%---------------------------------------
\item\label{ItemUseNorm} We can define graph rewriting systems over atomic flows that control normalisation procedures on derivations. This can be used to control a further refinement of the normalisation procedure (Theorem~\ref{ThNormAn}). 
\end{enumerate}
Our aim now is to quickly and informally provide the necessary notions about atomic flows, especially concerning aspects \eqref{ItemUseTop} and \eqref{ItemUseSubst} above. Although the feature \eqref{ItemUseNorm} of atomic flows, namely graph rewriting systems of flows, did help us in obtaining proofs in normal form, we estimate that formally introducing the necessary machinery is unjustified in this paper. In fact, given our limited needs here, we can operate directly on derivations, without the intermediate support of atomic flows. Nonetheless, being aware of the underlying atomic-flow methods is useful for the reader who wishes to further investigate this matter. So, we informally provide, in Section~\ref{SectNorm}, enough material to make the connection with the atomic-flow techniques that are fully developed in
\cite{GuglGund:07:Normalis:lr}.

We obtain one atomic flow from each derivation by tracing all its atom occurrences and by keeping track of their creation and destruction (in identity/cut and weak\-en\-ing/co\-weak\-en\-ing instances), their duplication (in contraction/cocontraction) and their duality (in identity/cut). Technically, atomic flows are directed graphs of a special kind, but it is more intuitive to consider them as diagrams generated by composing \emph{elementary atomic flows} that belong to one of seven kinds.

The first kind of elementary atomic flow is the \emph{edge}
\[
\vcenter{\hbox{\includegraphics{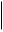}}}\quad,
\]
which corresponds to one or more occurrences of the same atom in a given derivation, all of which are not active in any structural rule instance, \emph{i.e.}, they are not the atom occurrences that instantiate a structural rule.

The other six kinds of elementary diagrams are associated with the six structural inference rules, as shown in Figure~\ref{FigVertAF}, and they are called \emph{vertices}; each vertex has some incident edges. At the left of each arrow, we see an instance of a structural rule, where the atom occurrences are labelled by small numerals; at the right of the arrow, we see the vertex corresponding to the rule instance, whose incident edges are labelled in accord with the atom occurrences they correspond to. We qualify each vertex according to the rule it corresponds to; for example, in a given atomic flow, we might talk about a \emph{contraction vertex}, or a \emph{cut vertex}, and so on. Instead of small numerals, sometimes we use $\epsilon$ or $\iota$ or colour to label edges (as well as atom occurrences), but we do not always use labels.

\newcommand{\one  }{{\mathchoice{\scriptstyle      \mathbf1}
                                {\scriptstyle      \mathbf1}
                                {\scriptstyle      \mathbf1}
                                {\scriptscriptstyle\mathbf1}}}
\newcommand{\two  }{{\mathchoice{\scriptstyle      \mathbf2}
                                {\scriptstyle      \mathbf2}
                                {\scriptstyle      \mathbf2}
                                {\scriptscriptstyle\mathbf2}}}
\newcommand{\three}{{\mathchoice{\scriptstyle      \mathbf3}
                                {\scriptstyle      \mathbf3}
                                {\scriptstyle      \mathbf3}
                                {\scriptscriptstyle\mathbf3}}}
%-------------------------------------------------------------------------------
\begin{figure}
\[
\begin{array}{@{}c@{}c@{}c@{}c@{}c@{}c@{}}
\vlinf\aid{}{\vls[a^\one.\bar a^\two]}\ttt&\quad{\to}\quad
   \vcenter{\hbox{\includegraphics{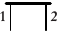}}}&\qquad\qquad
\vlinf\awd{}{a^\one}\fff&\quad{\to}\quad
   \vcenter{\hbox{\includegraphics{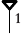}}}&\qquad\qquad
\vlinf\acd{}{a^\three}{\vls[a^\one.a^\two]}&\quad{\to}\quad
   \vcenter{\hbox{\includegraphics{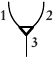}}}
\\
\noalign{\bigskip}
\vlinf\aiu{}\fff{\vls(a^\one.\bar a^\two)}&\quad{\to}\quad
   \vcenter{\hbox{\includegraphics{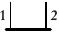}}}&\qquad\qquad
\vlinf\awu{}\ttt{a^\one}&\quad{\to}\quad
   \vcenter{\hbox{\includegraphics{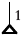}}}&\qquad\qquad
\vlinf\acu{}{\vls(a^\one.a^\two)}{a^\three}&\quad{\to}\quad
   \vcenter{\hbox{\includegraphics{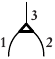}}}\\
\end{array}
\]
\caption{Vertices of atomic flows.}
\label{FigVertAF}
\end{figure}

All edges are directed, but we do not explicitly show the orientation. Instead, we consider it as implicitly given by the way we draw them, namely, edges are oriented along the vertical direction. So, the vertices corresponding to dual rules, in Figure~\ref{FigVertAF}, are distinct, for example, an identity vertex and a cut vertex are different because the orientation of their edges is different. On the other hand, the horizontal direction plays no role in distinguishing atomic flows; this corresponds to commutativity of logical relations.

\newcommand{\ppl}{{\mathchoice{\scriptstyle+}
                              {\scriptstyle+}
                              {\scriptstyle+}
                              {\scriptscriptstyle+}}}
\newcommand{\pmi}{{\mathchoice{\scriptstyle-}
                              {\scriptstyle-}
                              {\scriptstyle-}
                              {\scriptscriptstyle-}}}
We can define (\emph{atomic}) \emph{flows} as the smallest set of diagrams containing elementary atomic flows, and closed under the composition operation consisting in identifying zero or more edges such that no cycle is created. In addition, for a diagram to be an atomic flow, it must be possible to assign it a polarity, according to the following definition. A \emph{polarity assignment} is a mapping of each edge to an element of $\{\pmi,\ppl\}$, such that the two edges of each identity or cut vertex map to different values and the three edges of each contraction or cocontraction vertex map to the same value. We denote atomic flows by $\phi$ and $\psi$.

Let us see some examples. The flow
\begin{equation}\label{ExFlow}
\vcenter{\hbox{\includegraphics{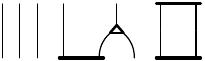}}}
\end{equation}
is obtained by juxtaposing (\emph{i.e.}, composing by identifying zero edges):
\begin{itemize}
%---------------------------------------
\item three edges, 
%---------------------------------------
\item a flow obtained by composing a cut vertex with a cocontraction vertex, and
%---------------------------------------
\item a flow obtained by composing an identity vertex with a cut vertex.
\end{itemize}
Note that there are no cycles in the flow, and that we can find 32 different polarity assignments, \emph{i.e.}, two for each of the five connected components of the flow (this is a general rule).

Let us see another example. These are three different representations of the same flow:
\[
\vcenter{\hbox{\includegraphics{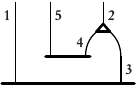}}}
\quad,\qquad
\vcenter{\hbox{\includegraphics{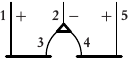}}}
\qquad\text{and}\qquad
\vcenter{\hbox{\includegraphics{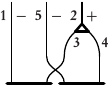}}}
\quad,
\]
where we label edges to show their correspondence. In the two rightmost flows, we indicate the two different polarity assignments that are possible.

The following two diagrams are not atomic flows:
\[
\vcenter{\hbox{\includegraphics{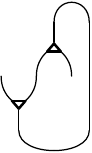}}}
\qquad\text{and}\qquad
\vcenter{\hbox{\includegraphics{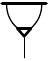}}}
\quad.
\]
The left one is not a flow because it contains a cycle, and the right one because there is no possible polarity assignment.

\newcommand{\four}{{\mathchoice{\scriptstyle      \mathbf4}
                               {\scriptstyle      \mathbf4}
                               {\scriptstyle      \mathbf4}
                               {\scriptscriptstyle\mathbf4}}}
\newcommand{\five}{{\mathchoice{\scriptstyle      \mathbf5}
                               {\scriptstyle      \mathbf5}
                               {\scriptstyle      \mathbf5}
                               {\scriptscriptstyle\mathbf5}}}
Let us see how to extract atomic flows from derivations. Given an $\SKS$ derivation $\Phi$, we obtain, by the following prescriptions, a unique atomic flow $\phi$, such that there is a surjective map between atom occurrences in $\Phi$ and edges of $\phi$:
\begin{itemize}
%-------------------
\item Each structural inference step in $\Phi$ is associated with one and only one vertex in $\phi$, such that active atom occurrences in the rule instance map to edges incident with the vertex. The correspondence is indicated in Figure~\ref{FigVertAF}. For example, the flow associated with the inference step at the left is indicated at the right:
\[
\vlinf\acd
      {}
      {\vls(a^\one.[b^\two.a^\five])}
      {\vls(a^\one.[b^\two.[a^\three.a^\four]])}
\qquad\text{and}\qquad
\vcenter{\hbox{\includegraphics{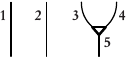}}}
\quad.
\]
Note that the nonactive atoms are `traced' by associating each trace with one edge; this corresponds well to abbreviating, say, the inference step $\vldownsmash{\vlinf\acd{}{K\vlscn[a]}{K\vlscn[a.a]}}$ by $K\left\{\vlinf{}{}a{\vls[a.a]}\right\}$.
%-------------------
\item For each other inference step in $\Phi$, all the atom occurrences in the premiss are respectively mapped to the same edges of $\phi$ as the atom occurrences in the conclusion. For example, the flow associated with the inference step
\[
\vlinf\med
      {}
      {\vls(a^\one.([b^\two.d^\four]).([c^\three.e^\five]))}
      {\vls(a^\one.[(b^\two.c^\three).(d^\four.e^\five)])}
\qquad\text{is}\qquad
\vcenter{\hbox{\includegraphics{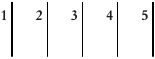}}}
\quad.
\]
\end{itemize}
The flow $\phi$ so obtained is called the atomic flow \emph{associated with} the derivation $\Phi$. We show three examples in Figure~\ref{FigExAF}: in the top row we see three $\SKS$ derivations in the standard CoS syntax; in the row below, we show the same derivations in the open deduction notation; in the bottom row, we see the three corresponding atomic flows.

\newcommand{\RD}[1]{{\color{Red}#1}}
\newcommand{\GR}[1]{{\color{Green}#1}}
\newcommand{\DO}[1]{{\color{DarkOrchid}#1}}
\newcommand{\PB}[1]{{\color{ProcessBlue}#1}}
\newcommand{\MG}[1]{{\color{Magenta}#1}}
\newcommand{\SG}[1]{{\color{SpringGreen}#1}}
\newcommand{\RS}[1]{{\color{RawSienna}#1}}
\newcommand{\YO}[1]{{\color{YellowOrange}#1}}
\newcommand{\PW}[1]{{\color{Periwinkle}#1}}
%-------------------------------------------------------------------------------
\begin{figure}
\[
%---------------------------------------
\begin{array}{@{}c@{}c@{}c@{}}
\vlderivation                                                {
\vlin=   {}{\ttt                                  }         {
\vlin\aiu{}{\vls[\fff.\ttt]                       }        {
\vlin=   {}{\vls[(\GR{a}.\RD{\bar a}).\ttt]       }       {
\vlin\swi{}{\vls[[(\RD{\bar a}.\GR{a}).\ttt].\ttt]}      {
\vlin=   {}{\vls[(\RD{\bar a}.[\GR{a}.\ttt]).\ttt]}     {
\vlin\swi{}{\vls[([\GR{a}.\ttt].\RD{\bar a}).\ttt]}    {
\vlin=   {}{\vls([\GR{a}.\ttt].[\RD{\bar a}.\ttt])}   {
\vlin\med{}{\vls([\GR{a}.\ttt].[\ttt.\RD{\bar a}])}  {
\vlin=   {}{\vls[(\GR{a}.\ttt).(\ttt.\RD{\bar a})]} {
\vlin\aid{}{\vls[\GR{a}.\RD{\bar a}]              }{
\vlhy      {\ttt                                  }}}}}}}}}}}}
\qquad&
%-------------------
\vlderivation                                                              {
\vlin\aiu{}
   {\vls(\DO{a}.\fff)                                            }        {
\vlin=   {}
   {\vls(\DO{a}.(\PB{a}.\MG{\bar a}))                            }       {
\vlin\acu{}
   {\vls((\DO{a}.\PB{a}).\MG{\bar a})                            }      {
\vlin=   {}
   {\vls(\SG{a}.\MG{\bar a})                                     }     {
\vlin\aiu{}
   {\vls([\fff.\SG{a}].\MG{\bar a})                              }    {
\vlin\acd{}
   {\vls([(\RD{a}.\RS{\bar a}).\SG{a}].\MG{\bar a})              }   {
\vlin\swi{}
   {\vls([(\RD{a}.[\GR{\bar a}.\YO{\bar a}]).\SG{a}].\MG{\bar a})}  {
\vlin=   {}
   {\vls((\RD{a}.[[\GR{\bar a}.\YO{\bar a}].\SG{a}]).\MG{\bar a})} {
\vlin\aid{}
   {\vls((\RD{a}.[\GR{\bar a}.[\YO{\bar a}.\SG{a}]]).\MG{\bar a})}{
\vlhy        
   {\vls((\RD{a}.[\GR{\bar a}.\ttt]).\MG{\bar a})                }}}}}}}}}}}
\qquad&
%-------------------
\vlderivation                                                            {
\vlin=   {}{\vls(([\RS{a}.\YO{b}].\PW{a}).([\GR{a}.\DO{b}].\SG{a}))}    {
\vlin\med{}{\vls(([\RS{a}.\YO{b}].[\GR{a}.\DO{b}]).(\PW{a}.\SG{a}))}   {
\vlin\acu{}{\vls([(\RS{a}.\GR{a}).(\YO{b}.\DO{b})].(\PW{a}.\SG{a}))}  {
\vlin\acu{}{\vls([(\RS{a}.\GR{a}).(\YO{b}.\DO{b})].\MG{a})         } {
\vlin\acu{}{\vls([(\RS{a}.\GR{a}).\PB{b}].\MG{a})                  }{
\vlhy      {\vls([\RD{a}.\PB{b}].\MG{a})                           }}}}}}}
\\
\noalign{\bigskip}
%---------------------------------------
\vlderivation                                                      {
\vlin\swi{}{\vlsbr[\vlinf{\swi}
                         {}
                         {\vls[\vlinf{}
                                     {}
                                     {\fff}
                                     {\vls(\GR{a}.\RD{\bar a})}
                              .
                               \ttt]}
                         {\vls([\GR{a}.\ttt].\RD{\bar a})}
                  \vlx.\vlx
                   \ttt
                  ]                                            }  {
\vlin\med{}{\vls([\GR{a}.\ttt].[\ttt.\RD{\bar a}])             } {
\vlin{}  {}{\vls[\GR{a}.\RD{\bar a}]                           }{
\vlhy      {\ttt                                               }}}}}
\qquad&
%-------------------
\vlinf=
      {}
      {\vls(\DO{a}.\vlinf{}{}\fff{\vls(\PB{a}.\MG{\bar a})})}      
      {\vlsbr(\vlinf\swi
                    {}
                    {\vls[\vlinf{}
                                {}
                                \fff
                                {\vls(\RD{a}
                                     .\vlinf{}
                                            {}
                                            {\RS{\bar a}}
                                            {\vls[\GR{\bar a}.\YO{\bar a}]}
                                     )}
                         \vlx.\vlx
                         \vlinf{}{}{\vls(\DO{a}.\PB{a})}{\SG{a}}
                         ]}
                    {\vls(\RD{a}
                         .[\GR{\bar a}
                          .\vlinf{}
                                 {}
                                 {\vls[\YO{\bar a}.\SG{a}]}
                                 \ttt
                          ]
                         )}
            \vlx.\vlx
            \MG{\bar a}
            )}
\qquad&
%-------------------
\vls(\vlinf\med
           {}
           {\vls([\RS{a}.\YO{b}].[\GR{a}.\DO{b}])}
           {\vls[\vlinf{}{}{\vls(\RS{a}.\GR{a})}{\RD{a}}
                .{\vlinf{}{}{\vls(\YO{b}.\DO{b})}{\PB{b}}}
                ]}
    \vlx.\vlx
     \vlinf{}{}{\vls(\PW{a}.\SG{a})}{\MG{a}}
    )
\\
\noalign{\bigskip}
%---------------------------------------
\vcenter{\hbox{\includegraphics{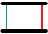}}}
\qquad&
%-------------------
\vcenter{\hbox{\includegraphics{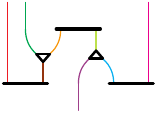}}}
\qquad&
%-------------------
\vcenter{\hbox{\includegraphics{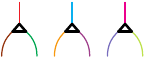}}}
\end{array}
\]
\caption{Examples of derivations in CoS and open deduction notation, and associated atomic flows.}
\label{FigExAF}
\end{figure}

Perhaps surprisingly, it can be proved that every flow is associated with infinitely many $\SKS$ derivations (see
\cite{GuglGund:07:Normalis:lr}).

We introduce now some graphical shortcuts. When certain details of a flow are not important, but only the vertex kinds and its upper and lower edges are, we can use boxes, labelled with all the vertex kinds that can appear in the flow they represent. For example, the following left and centre flows could represent the previously seen flow~\eqref{ExFlow}, whereas the right flow cannot:
\[
\vcenter{\hbox{\includegraphics{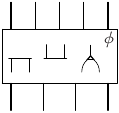}}}
\quad,\qquad
\vcenter{\hbox{\includegraphics{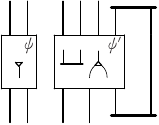}}}
\qquad\text{and}\qquad
\vcenter{\hbox{\includegraphics{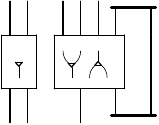}}}
\quad.
\]
The flow at the right cannot represent flow~\eqref{ExFlow} because it has the wrong number of lower edges and because a necessary cut vertex is not allowed by the labelling of the boxes. As just shown, we sometimes label boxes with the name of the flow they represent. For example, flow $\phi$ above could represent flow~\eqref{ExFlow}, and, if the centre flow stands for \eqref{ExFlow}, then flows $\psi$ and $\psi'$ are, respectively,
\[
\vcenter{\hbox{\includegraphics{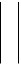}}}
\qquad\text{and}\qquad
\vcenter{\hbox{\includegraphics{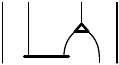}}}
\quad.
\]
When no vertex labels appear on a box, we assume that the vertices in the corresponding flow can be any (so, it does not mean that there are no vertices in the flow).

We sometimes use a double line notation for representing multiple edges. For example, the following diagrams represent the same flow:
\[
\vcenter{\hbox{\includegraphics{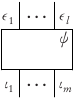}}}
\qquad\text{and}\qquad
\vcenter{\hbox{\includegraphics{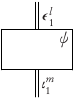}}}
\quad,
\]
where $l,m\ge0$; note that we use  $\boldsymbol\epsilon_1^l$ to denote the vector $(\epsilon_1,\dots,\epsilon_l)$. We might label multiple edges with one of the formulae that the associated atom occurrences form in a derivation.

We extend the double line notation to collections of isomorphic flows. For example, for $m\ge0$, the following diagrams represent the same flow:
\[
\vcenter{\hbox{\includegraphics{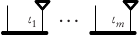}}}
\qquad\text{and}\qquad
\vcenter{\hbox{\includegraphics{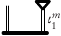}}}
\quad.
\]

We observe that the flow of every $\SKS$ derivation can always be represented as a collection of $m\ge0$ connected components as follows:
\[
\vcenter{\hbox{\includegraphics{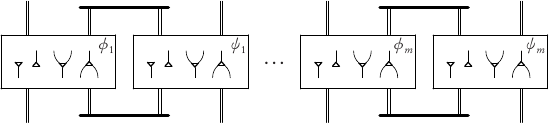}}}
\quad,
\]
such that each edge in flow $\phi_i$ is associated with some occurrence of some atom $a_i$, and each edge in flow $\psi_i$ is associated with some occurrence of atom $\bar a_i$. Note that it might happen that for $i\ne j$ we have $\vlsmash{a_i\equiv a_j}$. If we do not insist on dealing with connected components, we can adopt the same representation as above and stipulate that $i\ne j$ implies $\vlsmash{a_i\not\equiv a_j,\bar a_j}$. This would mean that the derivation only contains occurrences of atoms $a_1$, \dots, $a_m$, such that these atoms and their dual are all mutually distinct.

Note that no matter how we assign a polarity, all the edges in $\phi_i$ and all those in $\psi_i$ are respectively mapped to dual polarity values. Given a polarity assignment, we talk about \emph{negative} and \emph{positive} rule instances of (co)weakening and (co)contraction rules, according to whether the edges incident with the associated vertices map to $\pmi $ or $\ppl$, respectively.

In the following, when informally dealing with derivations, we freely transfer to them notions defined for their flows. For example, we can say that an atom occurrence is negative for a given polarity assignment (if the edge associated with the atom occurrence maps to $\pmi$) or that two atom occurrences are connected (if the associated edges belong to the same connected component). In fact, one of the advantages of working with flows is that they provide us with convenient geometrical notions.

As we mention at the beginning of this section, atomic flows help in selectively substituting for atom occurrences. In fact, given a derivation and its associated flow, we can use edges and boxes to individuate atom occurrences in the derivation, and then possibly substitute for them. For example, let us suppose that we are given the following associated derivation and flow:
\[
\Phi=
\vlsbr[\vlderivation                                              {
       \vlin{    }{}\fff                                         {
       \vlin{\med}{}{\vls(\vlinf{}{}a{\vls[a.a]}
                         .\vlinf={}{\bar a}{\vls[\fff.\bar a]})}{
       \vlhy        {\vls[(a.\fff).(a
                                   .\vlinf{}{}{\bar a}\fff)]   }}}}
      \vlx.\vlx\bar a]
\qquad\text{and}\qquad
\vcenter{\hbox{\includegraphics{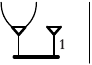}}}
\quad.
\]
We can then distinguish between the three occurrences of $\bar a$ that are mapped to edge $\one$ and the one that is not, as in
\[
\Phi=
\vlsbr[\vlderivation                                                        {
       \vlin{    }{}\fff                                                   {
       \vlin{\med}{}{\vls(\vlinf{}{}a{\vls[a.a]}
                         .\vlinf={}{\bar a^\one}{\vls[\fff.\bar a^\one]})}{
       \vlhy        {\vls[(a.\fff).(a
                                   .\vlinf{}{}{\bar a^\one}\fff)]        }}}}
      \vlx.\vlx\bar a]
\quad;
\]
we can also substitute for these occurrences, for example by $\{\bar a^\one/\fff\}$; such a situation occurs in the proof of Theorem~\ref{ThSimpleForm}. Note that simply substituting $\fff$ for $\bar a^\one$ would invalidate this derivation because it would break the cut and weakening instances; however, the proof of Theorem~\ref{ThSimpleForm} specifies how to fix such broken instances.

We generalise this labelling mechanism to boxes. For example, we can use a different representation of the flow of $\Phi$ to individuate two classes $a^\phi$ and $\bar a^\phi$ of atom occurrences, as follows:
\[
\Phi=
\vlsbr[\vlderivation                                                        {
       \vlin{    }{}\fff                                                   {
       \vlin{\med}{}{\vls(\vlinf{}{}{a^\phi}{\vls[a.a]}
                         .\vlinf={}{\bar a^\phi}{\vls[\fff.\bar a^\phi]})}{
       \vlhy        {\vls[(a.\fff).(a
                                   .\vlinf{}{}{\bar a^\phi}\fff)]        }}}}
      \vlx.\vlx\bar a^\phi]
\qquad\text{and}\qquad
\vcenter{\hbox{\includegraphics{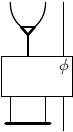}}}
\quad.
\]

In order to define the notion of cut-free form (Definition~\ref{DefNorm}), we need the following proposition, which we state here because it constitutes a good exercise about atomic flows. Note that, in the following, we use several boxes labelled by $\phi$: this means that we are dealing with several copies of the same flow $\phi$.

%-------------------------------------------------------------------------------
\begin{nota}\label{NotSubst}
Given a formula $A$ in a derivation whose associated atomic flow contains a flow $\phi$, we indicate with $a^\phi$ every occurrence of the atom $a$ in $A$ whose associated edge is in $\phi$. So, as in the following Proposition~\ref {PropSubst}, $A\{a^\phi/B,\bar a^\phi/\bar B\}$ stands for the formula $A$ where the atom occurrences of $a$ and its dual, whose associated edges are in $\phi$, are substituted with formula $B$ and its dual, respectively.
\end{nota}

%-------------------------------------------------------------------------------
\begin{prop}\label{PropSubst}
Given a derivation
\[
\vlder\Phi\SKS{A'}A
\quad,
\]
let its associated flow have shape
\[
\vcenter{\hbox{\includegraphics{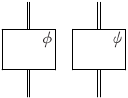}}}
\quad,
\]
such that $\phi$ is a connected component each of whose edges is associated with atom $a$ or $\bar a$; then, for any formula $B$, there exists a derivation
\[
\vlder\Psi\SKS{A'\{a^\phi/B,\bar a^\phi/\bar B\}}
              {A \{a^\phi/B,\bar a^\phi/\bar B\}}
\]
whose associated flow is
\[
\vcenter{\hbox{\includegraphics{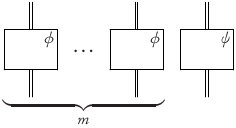}}}
\quad,
\]
where $m$ is the number of atom occurrences in $B$; moreover, the size of\/ $\Psi$ depends at most linearly on the size of\/ $\Phi$ and quadratically on the size of $B$.
\end{prop}

%-------------------------------------------------------------------------------
\begin{proof}
We can proceed by structural induction on $B$ and then on $\phi$. For the two cases of $B\equiv\vls[C.D]$ and $B\equiv\vls(C.D)$ we have to consider, for each vertex of $\phi$, one of the following situations:
\[%%%%%%%%%%%%%%%%%%%%%%%%% We shrank space: \quad -> \>\;\; and \qquad -> \quad
\vlderivation                                      {
\vlin\swi{}{\vls[C.D.(\bar C.\bar D)]           } { 
\vlin\swi{}{\vls[C.(\bar C.[D.\bar D])]         }{
\vlhy      {\vls(\vlinf{}{}{\vls[C.\bar C]}\ttt
                .\vlinf{}{}{\vls[D.\bar D]}\ttt)}}}}
\>\;\;,\quad
\vls[\vlinf{}{}C\fff.\vlinf{}{}D\fff]
\>\;\;,\quad
\vls(\vlinf{}{}C\fff.\vlinf{}{}D\fff)
\>\;\;,\quad
\vls[\vlinf{}{}C{\vls[C.C]}.\vlinf{}{}D{\vls[D.D]}]
\>\;\;,\quad
\vlinf\med
      {}
      {\vls(\vlinf{}{}C{\vls[C.C]}
           .\vlinf{}{}D{\vls[D.D]})}
      {\vls[(C.D).(C.D)]}
\>\;\;,
\]
and their dual ones.
\end{proof}

%-------------------------------------------------------------------------------
\begin{nota}\label{NotDerSubst}
In the hypotheses of Proposition~\ref{PropSubst}, we can describe $\Psi$ as $\Phi\{a^\phi/B,\bar a^\phi/\bar B\}$; one of $a^\phi/B$ or $\bar a^\phi/\bar B$ might be missing, when no identity or cut vertices are present in $\phi$.
\end{nota}

%===============================================================================
\section{Normalisation Step 1: Simple Form}\label{SectSimpleForm}

The first step in our normalisation procedure, defined here, consists in routine deep-inference manipulations, which are best understood in conjunction with atomic flows. For this reason, this section is a useful exercise for a reader who is not familiar with deep inference and atomic flows.

In this section, we define proofs in `simple form', in Definition~\ref{DefSimpleForm}, and we show that every proof can be transformed into simple form, in Theorem~\ref{ThSimpleForm}.

Let us establish the following conventions (they are especially useful to simplify our dealing with threshold formulae, in the next sections of the paper).

\newcommand{\avecletter}{{\boldsymbol a}}
\newcommand{\avec}[2]{\avecletter_{#1}^{#2}}
%-------------------------------------------------------------------------------
\begin{nota}
We use $\avec mn$ to denote the vector $(a_m,a_{m+1},\dots,a_n)$.
\end{nota}

%-------------------------------------------------------------------------------
\begin{conv}
When we talk about a set of \emph{distinct} atoms, we mean that no two atoms are the same or dual.
\end{conv}

%-------------------------------------------------------------------------------
\begin{defi}\label{DefSimpleForm}
Given a proof $\Pi$ of $A$ in $\SKS$, if there exist $n\ge0$ distinct atoms $a_1$, \dots, $a_n$ such that the proof and its atomic flow have shape, respectively,
\[
\hbox{\phantom{$\vls[A.{}]$}}
\vlderd\Psi
      {}
      {\vls[\llap{$\vls[A.{}]$}
            \vlinf{}{}\fff{\vls(a_1.\bar a_1^{\phi_1})}.\cdots.
            \vlinf{}{}\fff{\vls(a_n.\bar a_n^{\phi_n})}]}
      {\vls(\vlinf{}{}{\vls[a_1.\bar a_1 ^{\phi_1}]}\ttt.\cdots.
            \vlinf{}{}{\vls[a_n.\bar a_n^{\phi_n}]}\ttt)}
\qquad\text{and}\qquad
\vcenter{\hbox{\includegraphics{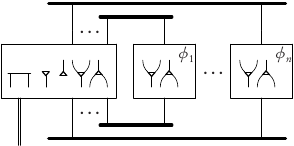}}}
\quad,
\]
we say that $\Pi$ is \emph{in simple form} (\emph{over\/ $\avec1n$}) and that $\Psi$ is a \emph{simple core} of $\Pi$.
\end{defi}

Proofs in simple form are such that all the cut instances are connected to identity instances via flows, the $\phi_i$ ones above, that only have one lower edge. The idea is that, in a proof in simple form, we can substitute formulae for all the occurrences of atoms $\bar a_i$ that map to some edge in $\phi_i$, without altering the conclusion of the proof. Of course, doing this would invalidate identity and cut instances, but we actually only need the simple core of the proof.

Our normalisation procedure essentially relies on gluing together simple cores, where we substitute the $a_i$ atom occurrences that map to edges in $\phi_i$ with certain formulae called `pseudocomplements' (see Section~\ref{SectThresh} and Definition~\ref{DefNorm}).

%-------------------------------------------------------------------------------
\begin{rem}
A proof in simple form over $\avec10$ is cut-free.
\end{rem}

In order to prove Theorem~\ref{ThSimpleForm}, we need two facts, Proposition~\ref{PropSwitch} and Lemma~\ref{LemmaCoContr}.

In the following (routine) proposition, we use the switch rule $\swi$ to `push outside' or `pull inside' a formula $A$, relative to a context $K\vlhole$.

%-------------------------------------------------------------------------------
\begin{prop}\label{PropSwitch}
For any context $K\vlhole$ and formula $A$, there exist derivations whose size is less than $\size{K\vlscn[A]}^2$ and have shape
\[
\vlder{}{\{\swi\}}{\vls[A.K\vlscn[\fff]]}{K\vlscn[A]}
\qquad\text{and}\qquad
\vlder{}{\{\swi\}}{K\vlscn(A)}{\vls(A.K\vlscn[\ttt])}
\quad.
\]
\end{prop}

%-------------------------------------------------------------------------------
\begin{proof}
We only build the derivation at the left in the claim, the construction being dual for the one at the right. We reason by induction on the number $n$ of $\vlor$-$\vlan$ alternations in the formula-tree branch of $\vlhole$ in $K\vlhole$. If $n=0$, then $K\vlscn[A]=\vls[A.K\vlscn[\fff]]$. If $n>0$, consider
\[
\vlsbr[\vlinf\swi
             {}
             {\vls[A.(H\vlscn[\fff].B)]}
             {\vlsbr(\vlder{}
                           {\{\swi\}}
                           {\vls[A.H\vlscn[\fff]]}
                           {H\vlscn[A]}
                    \vlx.\vlx B
                    )}
      \vlx.\vlx C]
\quad,
\]
for some context $H\vlhole$ and formulae $B$ and $C$, such that $\vls K\vlhole=[(H\vlhole.B).C]$ and the number of $\vlor$-$\vlan$ alternations in the formula-tree branch of $\vlhole$ in $H\vlhole$ is $n-1$. The number of $\swi$ instances is $n$, and we have that $n\le\size{ K\vlscn[\fff]}$.
\end{proof}

Note that the atomic flows of the derivations in the previous proposition only consist of edges because no structural rules appear.

To prove Theorem~\ref{ThSimpleForm}, we could now proceed as follows. Given a proof, we assign it (and its flow) an arbitrary polarity, under certain assumptions that we can always easily satisfy. We then focus on the negative paths connecting identity and cut vertices. If cocontraction vertices lie along these paths, we have a potential problem because some atoms in the conclusion of the proof might be connected to atoms in some identity instances. This would prevent us from substituting pseudocomplements, as previously mentioned, because by doing so we would alter the conclusion of the proof.

However, we can solve the problem by replacing each cocontraction vertex by an appropriate flow involving identity, cut and contraction vertices, in such a way that the only contraction vertex so introduced is positive. Actually, the lemma below takes a more radical approach, which simplifies exposition and also has broader application: we replace all negative contraction and cocontraction instances. This unnecessarily bloats the proof, but still stays well inside polynomial bounds.

%-------------------------------------------------------------------------------
\begin{lem}\label{LemmaCoContr}
Given any derivation
\[
\vlder\Phi\SKS BA
\quad,
\]
we can, in linear time in the size of\/ $\Phi$, construct a derivation
\[
\vlder{}\SKS BA
\]
such that its atomic flow has shape
\[
\vcenter{\hbox{\includegraphics{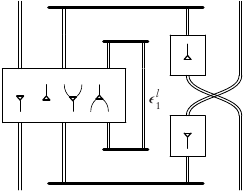}}}\quad,
\]
and such that no two atoms associated with $\epsilon_1$, \dots, $\epsilon_l$ are dual, for some $l\ge0$.
\end{lem}

%-------------------------------------------------------------------------------
\begin{proof}
Assign a polarity to the flow of $\Phi$ such that no two dual atoms are both associated with negative edges; then replace each negative contraction instance as follows:
\[
\vlderivation                           {
\vldd{\Psi'}\SKS B                  {
\vldd \Psi  \SKS{ K{\left\{
%-------------------
\vlinf{}
      {}
      {\bar a}
      {\vls[\bar a.\bar a]}
%-------------------
                            \right\}}}{
\vlhy           A                     }}}
\qquad\text{becomes}\qquad
\vlderivation                                                  {
\vldd{\Psi'}\SKS B                                            {
\vldd \Psi  \SKS{ K{\left\{
%-------------------
\vlinf\swi
      {}
      {\vlsbr[\vlderivation                             {
              \vlin\swi{}{\vls[\vlinf{}
                                     {}
                                     \fff
                                    {\vls(a.\bar a)}
                              .\vlinf{}
                                     {}
                                     \fff
                                     {\vls(a.\bar a)}
                              ]                      } {
              \vlin\swi{}{\vls(a.[(a.\bar a).\bar a])}{
              \vlhy      {\vls(\vlinf{}
                                     {}
                                     {\vls(a.a)}
                                     a          
                              .[\bar a. \bar a]
                              )                      }}}}
             \vlx.\vlx{\bar a}
             ]}
      {\vls(\vlinf{}{}{\vls[a.\bar a]}\ttt.[\bar a.\bar a])}
%-------------------
                     \right\}}                              }{
\vlhy           A                                            }}}
\quad.
\]
This corresponds, in the flow, to replacing each negative contraction vertex as follows:
\[
\vcenter{\hbox{\includegraphics{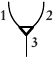}}}
\qquad\text{becomes}\qquad
\vcenter{\hbox{\includegraphics{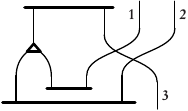}}}
\quad.
\]
Proceed analogously with negative cocontraction instances.
\end{proof}

We are now ready to prove the main result of this section.

%-------------------------------------------------------------------------------
\begin{thm}\label{ThSimpleForm}
Given any proof\/ $\Pi$ of $A$ in\/ $\SKS$, we can, in cubic time in the size of\/ $\Pi$, construct a proof of $A$ in simple form.
\end{thm}

\newcommand{\bvecletter}{{\boldsymbol b}}
\newcommand{\bvec}[2]{\bvecletter_{#1}^{#2}}
\newcommand{\Ord}[1]{{\mathsf O}(#1)}
%-------------------------------------------------------------------------------
\begin{proof}
We proceed in three steps.
\begin{enumerate}
%---------------------------------------
\item\label{ItemWeak} By Lemma~\ref{LemmaCoContr}, we can transform $\Pi$, in linear time in its size, into a proof $\vlsmash{\vlproof{\Pi'}\SKS A}$, whose flow has shape
\[
\vcenter{\hbox{\includegraphics{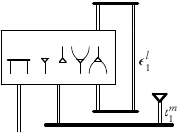}}}\quad,
\]
where $l,m\ge0$ and such that no two atoms associated with $\epsilon_1$, \dots, $\epsilon_l$ are dual. For $1\le i\le m$, we successively transform $\Pi'$ as follows, for some $\Pi''$, $\Phi$, $\Phi'$, $ K\vlhole$ and $H\vlhole$:
\[
\vlderivation                                                       {
\vldd{\Phi'}{}A                                                    {
\vldd{\Phi }{}{H{\left\{\vlinf{}
                              {}
                              \fff
                              {\vls(a.\bar a^{\iota_i})}\right\}}}{
\vlpd{\Pi''} {}{K{\left\{\vlinf{}
                               {}
                               {\bar a^{\iota_i}}
                               \fff              \right\}}       }}}}
\qquad\text{becomes}\qquad
\vlderivation                                       {
\vldd{\Phi'}{}A                                    {
\vldd{\Phi\{\bar a^{\iota_i}/\fff\}
           }{}{H{\left\{\vls(\vlinf{}
                                   {}
                                   \ttt
                                   a
                                  .\fff)\right\}}}{
\vlpr{\Pi''}{}{K\vlscn[\fff]                     }}}}
\quad.
\]
This way, we obtain, in linear time, a proof $\vlproof{\Pi'''}\SKS A$, whose flow is
\[
\vcenter{\hbox{\includegraphics{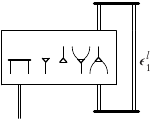}}}
\quad,
\]
and whose size is smaller than $\size{\Pi'}$.
%---------------------------------------
\item Thanks to Proposition~\ref{PropSwitch}, for $1\le i\le l$, we successively transform $\Pi'''$ as follows, for some $\Psi$, $\Psi'$ and $K'\vlhole$:
\[
\vlderivation                                                {
\vldd{\Psi'}{}A                                             {
\vlpd{\Psi }{}{K'{\left\{\vlinf{}
                               {}
                               {\vls[a.\bar a^{\epsilon_i}]}
                               {\vls[\ttt]}\right\}}       }}}
\qquad\text{becomes}\qquad
\vlderivation                                                       {
\vlde{\Psi'}{}A                                                    {
\vlde{}     {\{\swi\}}
              {      K'\vlscn       [a.\bar a^{\epsilon_i}]    }  {
\vlde{\vls([a.\bar a].\Psi)}
            {}{\vls([a.\bar a^{\epsilon_i}].{K'\vlscn[\ttt]})  } {
\vlin{}     {}{\vls[a.\bar a^{\epsilon_i}]                     }{
\vlhy                \ttt                                       }}}}}
\quad;
\]
we also apply the dual transformation for each $\aiu$ instance. This way, we obtain a proof
\[
\hbox{\phantom{$\vls[A.{}]$}}
\vlderd{\Psi''}
       {}
       {\vls[\llap{$\vls[A.{}]$}
             \vlinf{}{}\fff{\vls(a_1.\bar a_1^{\epsilon_1})}.\cdots.
             \vlinf{}{}\fff{\vls(a_l.\bar a_l^{\epsilon_l})}]}
       {\vls(\vlinf{}{}{\vls[a_1.\bar a_1^{\epsilon_1}]}\ttt.\cdots.
             \vlinf{}{}{\vls[a_l.\bar a_l^{\epsilon_l}]}\ttt)}
\quad,
\]
whose flow is the same as that of $\Pi'''$ because each transformation conserves the flow. If $\size{\Pi'''}=n$, and given that $n>2l$, the size of each derivation introduced by virtue of Proposition~\ref{PropSwitch} is at most $4n^2$. So, each of the $2l$ transformations increases the size of the proof by $\Ord{n^2}$, which makes for a total complexity of $\Ord{n^3}$.
%---------------------------------------
\item Consider $\bvec1n$ such that $b_1$, \dots, $b_n$ are distinct and $\{a_1,\dots,a_l\}=\{b_1,\dots,b_n\}$. We can build, in linear time, the proof
\[ %%%%% The \dimen's must be adjusted if fonts and layout parameters are changed
\dimen0=2420000sp
\kern\dimen0
\vlderivation                                                        {
\vldd{\Psi''}{}{\kern-\dimen0\vlsbr[A\vlx.\vlx\vlderd
                {}
                {\{\gcd\}}
                {\vlsbr[\vlinf{}{}\fff{\vls(b_1.\bar b_1)}.\cdots.
                        \vlinf{}{}\fff{\vls(b_n.\bar b_n)}]}
                {\vlsmallbrackets
                 \vls[(a_1.\bar a_1^{\epsilon_1}).\cdots.
                      (a_l.\bar a_l^{\epsilon_l})]}]              } {
\vldd{      }{\{\gcu\}
              }{\vlsmallbrackets
                \vls([a_1.\bar a_1^{\epsilon_1}].\cdots.
                     [a_l.\bar a_l^{\epsilon_l}])                 }{
\vlhy        {\vls(\vlinf{}{}{\vls[b_1.\bar b_1]}\ttt.\cdots.
                 \vlinf{}{}{\vls[b_n.\bar b_n]}\ttt)              }}}}
\quad,
\]
which is in simple form over $\bvec1n$. We can then obtain a proof in $\SKS$ in time $\Ord{n^2}$, because of Proposition~\ref{PropGenAtPol}. 
\end{enumerate}
\vskip-\baselineskip
\end{proof}

\noindent The transformation in Step~\eqref{ItemWeak} in the previous proof is a case of `weakening reduction' for atomic flows, studied in
\cite{GuglGund:07:Normalis:lr}. In Section~\ref{SectNorm} we comment more on this.

%-------------------------------------------------------------------------------
\begin{rem}
In general, given a proof $\Pi$ and by the construction in the proof of Theorem~\ref{ThSimpleForm}, we can obtain several different simple forms from $\Pi$. In fact, apart from permutations of rule instances, commutativity and associativity, the simple forms depend on the choice of a polarity assignment (Lemma~\ref{LemmaCoContr}).
\end{rem}

%===============================================================================
\section{Threshold Formulae}\label{SectThresh}

\newcommand{\Gammasf}{\mathsf\Gamma}
We present here the main construction of this paper, \emph{i.e.}, a class of derivations $\Gammasf$ that only depend on a given set of atoms and that allow us to normalise any proof containing those atoms. The complexity of the $\Gammasf$ derivations dominates the complexity of the normal proof, and is due to the complexity of certain `threshold formulae', on which the $\Gammasf$ derivations are based. The $\Gammasf$ derivations are constructed in Definition~\ref{DefThrDer}; this directly leads to Theorem~\ref{TheoThrDer}, which states a crucial property of the $\Gammasf$ derivations and which is the main result of this section.

Threshold formulae realise boolean threshold functions, which are defined as boolean functions that are true if and only if at least $k$ of $n$ inputs are true (see
\cite{Wege:87:The-Comp:vn} for a thorough reference on threshold functions). 

In the following, $\lfloor x\rfloor$ denotes the maximum integer $n$ such that $n\le x$.

There are several ways of encoding threshold functions into formulae, and the problem is to find, among them, an encoding that allows us to obtain Theorem~\ref{TheoThrDer}. Efficiently obtaining the property stated in Theorem~\ref{TheoThrDer} crucially depends also on the proof system we adopt.

The following class of threshold formulae, which we found to work for system $\SKS$, is a simplification of the one adopted in
\cite{AtseGalePudl:02:Monotone:yu}.

\renewcommand{\th}[2]{\mathop{\uptheta_{#1}^{#2}}}
%-------------------------------------------------------------------------------
\begin{defi}
Consider $n>0$, distinct atoms $a_1$, \dots, $a_n$, and let $p=\lfloor n/2\rfloor$ and $q=n-p$; for $k\ge0$, we define the \emph{threshold formulae\/} $\th kn\avec1n$ as follows:
\begin{itemize}
%---------------------------------------
\item for any $n>0$ let $\th0n\avec1n\equiv\ttt$;
%---------------------------------------
\item for any $n>0$ and $k>n$ let $\th kn\avec1n\equiv\fff$;
%---------------------------------------
\item $\th11(a_1)\equiv a_1$;
%---------------------------------------
\item for any $n>1$ and $0<k\le n$ let
$\th kn\avec1n\equiv\bigvee_{\begin{subarray}{l}i+j=k      \\ 
                                                0\le i\le p\\ 
                                                0\le j\le q
                             \end{subarray}}
\vlsbr(\th ip\avec1p.\th jq\avec{p+1}n)$.
\end{itemize}
\end{defi}

Compared to the definition in \cite{AtseGalePudl:02:Monotone:yu}, we require $i+j=k$ instead of $i+j\ge k$. The semantics of the formulae does not change but their size is smaller and their structure is much simpler, arguably benefiting further research. See, in Figure~\ref{FigThrEx}, some examples of threshold formulae.

The only reason why we require atoms to be distinct in threshold formulae is to avoid certain technical problems with substitutions in the definition of cut-free form, later on. However, there is no substantial difficulty in relaxing this definition to any set of atoms.

%-------------------------------------------------------------------------------
\begin{figure}
\vlsmallbrackets
\begin{eqnarray*}
%---------------------------------------
\th02(a,b)&\equiv&\ttt
\quad,\\
\noalign{\medskip}
%---------------------------------------
\th12(a,b)&\equiv&\vls[({\vlnos\th11(a)}.{\vlnos\th01(b)}).
                       ({\vlnos\th01(a)}.{\vlnos\th11(b)})]
           \equiv     [(a.\ttt).(\ttt.b)]\\
          &=     &\vls [a      .      b ]
\quad,\\
\noalign{\medskip}
%---------------------------------------
\th22(a,b)&\equiv&\vls({\vlnos\th11(a)}.{\vlnos\th11(b)})\\
          &\equiv&\vls(a.b)
\quad,\\
\noalign{\medskip}
%---------------------------------------
\th03(a,b,c)&\equiv&\ttt
\quad,\\
\noalign{\medskip}
%---------------------------------------
\th13(a,b,c)&\equiv&\vls[({\vlnos\th11(a)}.{\vlnos\th02(b,c)}).
                         ({\vlnos\th01(a)}.{\vlnos\th12(b,c)})]
             \equiv     [(a.\ttt).(\ttt.[(b.\ttt).(\ttt.c)])]\\
            &=     &\vls[a.b.c]
\quad,\\
\noalign{\medskip}
%---------------------------------------
\th23(a,b,c)&\equiv&\vls[({\vlnos\th11(a)}.{\vlnos\th12(b,c)}).
                    ({\vlnos\th01(a)}.{\vlnos\th22(b,c)})]\\
            &=     &\vls[(a.[b.c]).(b.c)]
\quad,\\
\noalign{\medskip}
%---------------------------------------
\th33(a,b,c)&\equiv&\vls({\vlnos\th11(a)}.{\vlnos\th22(b,c)})
             \equiv     [(a.(b.c))]\\
            &=     &\vls(a.b.c)
\quad,\\
\noalign{\medskip}
%---------------------------------------
\th05(a,b,c,d,e)&\equiv&\ttt
\quad,\\
\noalign{\medskip}
%---------------------------------------
\th15(a,b,c,d,e)&\equiv&\vls[({\vlnos\th12(a,b)}.{\vlnos\th03(c,d,e)}).
                             ({\vlnos\th02(a,b)}.{\vlnos\th13(c,d,e)})]\\
                &=     &\vls[a.b.c.d.e]
\quad,\\
\noalign{\medskip}
%---------------------------------------
\th25(a,b,c,d,e)&\equiv&\vls[({\vlnos\th22(a,b)}.{\vlnos\th03(c,d,e)}).
                             ({\vlnos\th12(a,b)}.{\vlnos\th13(c,d,e)}).
                             ({\vlnos\th02(a,b)}.{\vlnos\th23(c,d,e)})]\\
                &=     &\vls[(a.b                                    ).
                             ([a.b]             .[c.d.e]             ).
                                                 (c.[d.e]).(d.e)      ]
\quad,\\
\noalign{\medskip}
%---------------------------------------
\th35(a,b,c,d,e)&\equiv&\vls[({\vlnos\th22(a,b)}.{\vlnos\th13(c,d,e)}).
                             ({\vlnos\th12(a,b)}.{\vlnos\th23(c,d,e)}).
                             ({\vlnos\th02(a,b)}.{\vlnos\th33(c,d,e)})]\\
                &=     &\vls[(a.b               .[c.d.e]             ).
                             ([a.b]             .[(c.[d.e]).(d.e)]   ).
                                                 (c.d.e)              ]
\quad,\\
\noalign{\medskip}
%---------------------------------------
\th45(a,b,c,d,e)&\equiv&\vls[({\vlnos\th22(a,b)}.{\vlnos\th23(c,d,e)}).
                             ({\vlnos\th12(a,b)}.{\vlnos\th33(c,d,e)})]\\
                &=     &\vls[(a.b               .[(c.[d.e]).(d.e)]   ).
                             ([a.b]             .c.d.e               )]
\quad,\\
\noalign{\medskip}
%---------------------------------------
\th55(a,b,c,d,e)&\equiv&\vls({\vlnos\th22(a,b)}.{\vlnos\th33(c,d,e)})\\
                &=     &\vls(a.b.c.d.e)
\quad,\\
\noalign{\medskip}
%---------------------------------------
\th65(a,b,c,d,e)&\equiv&\fff
\quad.
\end{eqnarray*}
\caption{Examples of threshold formulae.}
\label{FigThrEx}
\end{figure}

The formulae for threshold functions adopted in
\cite{AtseGalePudl:02:Monotone:yu} correspond, for each choice of $k$ and $n$, to $\bigvee_{i\ge k}\th in\avec1n$. We presume that
\cite{AtseGalePudl:02:Monotone:yu} employs these more complicated formulae because the formalism adopted there, the sequent calculus, is less flexible than deep inference, requiring more information in threshold formulae in order to construct suitable derivations.

%-------------------------------------------------------------------------------
\begin{rem}
For $n>0$, we have $\th1n\avec1n=\vls[a_1.\vldots.a_n]$ and $\th nn\avec1n=\vls(a_1.\vldots.a_n)$.
\end{rem}

The size of the threshold formulae dominates the cost of the normalisation procedure, so, we evaluate their size. We leave as an exercise the proof of the following proposition.

%-------------------------------------------------------------------------------
\begin{prop}\label{PropQuasAux}
For any $n>0$ and $k\ge0$, $\size{\th kn\avec1n}\le\size{\th{\lfloor n/2\rfloor+1}n\avec1n}$.
\end{prop}

%-------------------------------------------------------------------------------
\begin{lem}\label{LemmaQuas}
The size of\/ $\th{\lfloor n/2\rfloor+1}n\avec1n$ is $n^{\Ord{\log n}}$.
\end{lem}

%-------------------------------------------------------------------------------
\begin{proof}
Observe that $\size{\th kn\avec1n}\le\size{\th k{n+1}\avec1{n+1}}$. Let $p=\lfloor n/2\rfloor$ and $q=n-p$ and consider:
\begin{equation}\label{PropQuasIneq}
\begin{split}
\size{\th{p+1}n\avec1n}
&=\textstyle\sum_{\begin{subarray}{l}i+j=p+1    \\
                                     0\le i\le p\\
                                     0\le j\le q
                  \end{subarray}}
  \left(\size{\th ip\avec1p}+
        \size{\th jq\avec{p+1}n}\right)             \\
&\le\textstyle\sum_{\begin{subarray}{l}i+j=p+1\\
                                       0\le i,j\le q
                    \end{subarray}}
  \left(\size{\th iq\avec1q}+
        \size{\th jq\avec1q}\right)                 \\
&\le2(q+1)
  \size{\th{\lfloor q/2\rfloor+1}q\avec1q}\quad,
\end{split}
\end{equation}
where we use Proposition~\ref{PropQuasAux}. We show that, for $h=2/(\log3-\log2)$ and for any $n>0$, we have $\size{\th{\lfloor n/2\rfloor+1}n\avec1n}\le n^{h\log n}$. We reason by induction on $n$; the case $n=1$ trivially holds. By the inequality~\eqref{PropQuasIneq}, and for $n>1$, we have
\begin{equation*}
\begin{split}
\size{\th{\lfloor n/2\rfloor+1}n\avec1n}
&\le2(n-\lfloor n/2\rfloor+1)
     (n-\lfloor n/2\rfloor)^{h\log(n-\lfloor n/2\rfloor)}       \\
&\le n^2n^{h\log(2n/3)}=n^{h\log n-h(\log3-\log2)+2}=n^{h\log n}
\quad.
\end{split}
\end{equation*}
\vskip-\baselineskip
\end{proof}
\newpage
%-------------------------------------------------------------------------------
\begin{thm}\label{TheoQuas}
For any $k\ge0$ the size of\/ $\th kn\avec1n$ is $n^{\Ord{\log n}}$.
\end{thm}

%-------------------------------------------------------------------------------
\begin{proof}
It immediately follows from Proposition~\ref{PropQuasAux} and Lemma~\ref{LemmaQuas}.
\end{proof}

Given a threshold formula $\th kn\avec1n$, we can consider, for each $a_l$ such that $1\le l\le n$, the formulae $(\th kn\avec1n)\{a_l/\fff\}$ and $(\th{k+1}n\avec1n)\{a_l/\ttt\}$: we call both of them, informally, `pseudocomplements' of $a_l$. The reason for this name is that we can manage to replace, in a given proof, all occurrences of those $\bar a_l$ that appear in cut instances with the pseudocomplements of $a_l$. The cut instances and their corresponding identity instances are then removed, leaving us with derivations whose premiss and conclusion contain each a threshold formula. Moreover, the $k$-level of the threshold formula in the premiss is one less than the $k$-level of the threshold formula in the conclusion. This way, we obtain several derivations, corresponding to increasing values of $k$, that we are able to stitch together until we get a normalised proof.

All this, of course, needs clarification and for many it might be helpful only after having grasped the full proof in its technical form. However, we think that it is convenient here to provide a summary of the main constructions allowing for this stitching operation. Let us read derivations top-down; the following are the steps that we need to perform, for $0\le k\le n$.
\begin{enumerate}
%---------------------------------------
\item\label{ItemOne} Build
\[
\vlder{}{}{\vlsmallbrackets\vls[a_l.(\th kn\avec1n)\{a_l/\fff\}]}
          {\th kn\avec1n}
\quad,
\]
\emph{i.e.}, create, from a $k$-level threshold formula, a disjunction between $a_l$ and its pseudocomplement $(\th kn\avec1n)\{a_l/\fff\}$ (Proposition~\ref{PropAuxNorm}); then replace the pseudocomplement into $\bar a_l$, for each identity instance.
%---------------------------------------
\item\label{ItemTwo} Increase the $k$-level by using the derivations
\[
\vlder{}{}{(\th{k+1}n\avec1n)\{a_l/\ttt\}}
          {(\th kn\avec1n)\{a_l/\fff\}}
\]
(Theorem~\ref{TheoThrDer}); these are the $\Gammasf$ derivations mentioned in the introduction to this section.
%---------------------------------------
\item\label{ItemThree} For each cut instance, collect the conjunction between $a_l$ and its pseudocomplement $(\th{k+1}n\avec1n)\{a_l/\ttt\}$; then build
\[
\vlder{}{}{\th{k+1}n\avec1n}
          {\vlsmallbrackets\vls(a_l.(\th{k+1}n\avec1n)\{a_l/\ttt\})}
\quad,
\]
\emph{i.e.}, create a $(k+1)$-level threshold formula (Proposition~\ref{PropAuxNorm}).
\end{enumerate}
The derivations mentioned above do not require any use of identity and cut, and allow us to move, in $n+1$ steps, from $\th 0n\avec1n\equiv\ttt$ to $\th{n+1}n\avec1n\equiv\fff$, which is the secret to success. The constructions in~\ref{ItemOne} and \ref{ItemThree} are deep-inference routine and introduce low complexity. We deal now with the crucial step~\ref{ItemTwo}, by designing Definition~\ref{DefThrDer}, and then checking it carefully, so as to get the property stated in Theorem~\ref{TheoThrDer}.

Definition~\ref{DefThrDer} is technical, but its philosophy is simple; all one has to do to build the derivations required by Theorem~\ref{TheoThrDer} is:
\begin{itemize}
%---------------------------------------
\item identify the atom occurrences that must occur in the premiss and that must not occur in the conclusion and remove them using coweakening, and
%---------------------------------------
\item identify the atom occurrences that must occur in the conclusion and that must not occur in the premiss and add them using weakening.
\end{itemize}
We have implemented Definition~\ref{DefThrDer} as a program
\cite{Gugl:09:th.pl:rz}. It can be useful to read the definition together with the examples in Figures~\ref{FigPThEx} and \ref{FigThrEx}, which have been generated by the program.

\newcommand{\Gth}[3]{\mathop{\Gammasf_{#1,#2}^{#3}}}
%-------------------------------------------------------------------------------
\begin{figure}
\begingroup
\footnotesize
\begin{eqnarray*}
%---------------------------------------
\Gth 015\avecletter&=&
\vls [\ttt.\vlderivation{
\vlin{}{}{b}{
\vlhy{\vls \fff}
}}
.\vlderivation{
\vlin{}{}{\vls [c.d.e]}{
\vlhy{\vls \fff}
}}
]\quad,\\
\noalign{\smallskip}
%---------------------------------------
\Gth 115\avecletter&=&
\vls [b.([\ttt.\vlderivation{
\vlin{}{}{b}{
\vlhy{\vls \fff}
}}
].[c.d.e]).\vlderivation{
\vlin{}{}{\vls [(c.[d.e]).(d.e)]}{
\vlhy{\vls \fff}
}}
]\quad,\\
\noalign{\smallskip}
%---------------------------------------
\Gth 215\avecletter&=&
\vls [(b.[c.d.e]).([\ttt.\vlderivation{
\vlin{}{}{b}{
\vlhy{\vls \fff}
}}
].[(c.[d.e]).(d.e)]).\vlderivation{
\vlin{}{}{\vls \fff}{
\vlhy{\vls (\fff.b)}
}}
.\vlderivation{
\vlin{}{}{\vls (c.d.e)}{
\vlhy{\vls \fff}
}}
]\quad,\\
\noalign{\smallskip}
%---------------------------------------
\Gth 315\avecletter&=&
\vls [(b.[(c.[d.e]).(d.e)]).([\ttt.\vlderivation{
\vlin{}{}{b}{
\vlhy{\vls \fff}
}}
].c.d.e).\vlderivation{
\vlin{}{}{\vls \fff}{
\vlhy{\vls (\fff.b.[c.d.e])}
}}
]\quad,\\
\noalign{\smallskip}
%---------------------------------------
\Gth 415\avecletter&=&
\vls [(b.c.d.e).\vlderivation{
\vlin{}{}{\vls \fff}{
\vlhy{\vls (\fff.b.[(c.[d.e]).(d.e)])}
}}
]\quad,\\
\noalign{\smallskip}
%---------------------------------------
\Gth 515\avecletter&=&
\vlderivation{
\vlin{}{}{\vls \fff}{
\vlhy{\vls (\fff.b.c.d.e)}
}}
\quad,\\
\noalign{\smallskip}
%---------------------------------------
\Gth 035\avecletter&=&
\vls [\ttt.\vlderivation{
\vlin{}{}{\vls [d.e]}{
\vlhy{\vls \fff}
}}
.\vlderivation{
\vlin{}{}{\vls [a.b]}{
\vlhy{\vls \fff}
}}
]\quad,\\
\noalign{\smallskip}
%---------------------------------------
\Gth 135\avecletter&=&
\vls [([a.b].[\ttt.\vlderivation{
\vlin{}{}{\vls [d.e]}{
\vlhy{\vls \fff}
}}
]).d.e.\vlderivation{
\vlin{}{}{\vls (d.e)}{
\vlhy{\vls \fff}
}}
.\vlderivation{
\vlin{}{}{\vls (a.b)}{
\vlhy{\vls \fff}
}}
]\quad,\\
\noalign{\smallskip}
%---------------------------------------
\Gth 235\avecletter&=&
\vls [(a.b.[\ttt.\vlderivation{
\vlin{}{}{\vls [d.e]}{
\vlhy{\vls \fff}
}}
]).([a.b].[d.e.\vlderivation{
\vlin{}{}{\vls (d.e)}{
\vlhy{\vls \fff}
}}
]).(d.e).\vlderivation{
\vlin{}{}{\vls \fff}{
\vlhy{\vls (\fff.[d.e])}
}}
]\quad,\\
\noalign{\smallskip}
%---------------------------------------
\Gth 335\avecletter&=&
\vls [(a.b.[d.e.\vlderivation{
\vlin{}{}{\vls (d.e)}{
\vlhy{\vls \fff}
}}
]).([a.b].[(d.e).\vlderivation{
\vlin{}{}{\vls \fff}{
\vlhy{\vls (\fff.[d.e])}
}}
]).\vlderivation{
\vlin{}{}{\vls \fff}{
\vlhy{\vls (\fff.d.e)}
}}
]\quad,\\
\noalign{\smallskip}
%---------------------------------------
\Gth 435\avecletter&=&
\vls [(a.b.[(d.e).\vlderivation{
\vlin{}{}{\vls \fff}{
\vlhy{\vls (\fff.[d.e])}
}}
]).\vlderivation{
\vlin{}{}{\vls \fff}{
\vlhy{\vls ([a.b].\fff.d.e)}
}}
]\quad,\\
\noalign{\smallskip}
%---------------------------------------
\Gth 535\avecletter&=&
\vlderivation{
\vlin{}{}{\vls \fff}{
\vlhy{\vls (a.b.\fff.d.e)}
}}\quad,\\
\noalign{\smallskip}
%---------------------------------------
\Gth 055\avecletter&=&
\vls [\ttt.\vlderivation{
\vlin{}{}{d}{
\vlhy{\vls \fff}
}}
.\vlderivation{
\vlin{}{}{c}{
\vlhy{\vls \fff}
}}
.\vlderivation{
\vlin{}{}{\vls [a.b]}{
\vlhy{\vls \fff}
}}
]\quad,\\
\noalign{\smallskip}
%---------------------------------------
\Gth 155\avecletter&=&
\vls [([a.b].[\ttt.\vlderivation{
\vlin{}{}{d}{
\vlhy{\vls \fff}
}}
.\vlderivation{
\vlin{}{}{c}{
\vlhy{\vls \fff}
}}
]).(c.[\ttt.\vlderivation{
\vlin{}{}{d}{
\vlhy{\vls \fff}
}}
]).d.\vlderivation{
\vlin{}{}{\vls (a.b)}{
\vlhy{\vls \fff}
}}
]\quad,\\
\noalign{\smallskip}
%---------------------------------------
\Gth 255\avecletter&=&
\vls [(a.b.[\ttt.\vlderivation{
\vlin{}{}{d}{
\vlhy{\vls \fff}
}}
.\vlderivation{
\vlin{}{}{c}{
\vlhy{\vls \fff}
}}
]).([a.b].[(c.[\ttt.\vlderivation{
\vlin{}{}{d}{
\vlhy{\vls \fff}
}}
]).d]).(c.d).\vlderivation{
\vlin{}{}{\vls \fff}{
\vlhy{\vls (d.\fff)}
}}
]\quad,\\
\noalign{\smallskip}
%---------------------------------------
\Gth 355\avecletter&=&
\vls [(a.b.[(c.[\ttt.\vlderivation{
\vlin{}{}{d}{
\vlhy{\vls \fff}
}}
]).d]).([a.b].[(c.d).\vlderivation{
\vlin{}{}{\vls \fff}{
\vlhy{\vls (d.\fff)}
}}
]).\vlderivation{
\vlin{}{}{\vls \fff}{
\vlhy{\vls (c.d.\fff)}
}}
]\quad,\\
\noalign{\smallskip}
%---------------------------------------
\Gth 455\avecletter&=&
\vls [(a.b.[(c.d).\vlderivation{
\vlin{}{}{\vls \fff}{
\vlhy{\vls (d.\fff)}
}}
]).\vlderivation{
\vlin{}{}{\vls \fff}{
\vlhy{\vls ([a.b].c.d.\fff)}
}}
]\quad,\\
\noalign{\smallskip}
%---------------------------------------
\Gth 555\avecletter&=&
\vlderivation{
\vlin{}{}{\vls \fff}{
\vlhy{\vls (a.b.c.d.\fff)}
}}\quad.
\end{eqnarray*}
\endgroup
\caption{Examples of $\Gth kl5\avecletter$, where $\avecletter=(a,b,c,d,e)$.}
\label{FigPThEx}
\end{figure}

%-------------------------------------------------------------------------------
\begin{rem}
Given $n>1$, let $p=\lfloor n/2\rfloor$ and $q=n-p$. For $0\le k\le q$ and $1\le l\le p$, the following derivation is well defined:
\[
\vlinf{\gwu}
      {}
      {\fff}
      {\vls({\vlnos(\th pp\avec1p)}\{a_l/\fff\}.\th kq\avec{p+1}n)}
=
\vls(
\vlinf{\gwu}
      {}
      {\vls(\ttt)}
      {\vls(a_1.\cdots.a_{l-1}.a_{l+1}.\cdots.a_p.\th kq\avec{p+1}n)}
.\fff)
\quad.
\]
Analogously, for $0\le k\le p$ and $p+1\le l\le n$, we can define the following derivation:
\[
\vlinf{\gwu}
      {}
      {\fff}
      {\vls(\th kp\avec1p.{\vlnos(\th qq\avec{p+1}n)}\{a_l/\fff\})}
=
\vls(
\vlinf{\gwu}
      {}
      {\vls(\ttt)}
      {\vls(\th kp\avec1p.a_{p+1}.\cdots.a_{l-1}.a_{l+1}.\cdots.a_n)}
.\fff)
\quad.
\]
Both classes of derivations are used in Definition~\ref{DefThrDer}.
\end{rem}

\newcommand{\Uth}[3]{\mathop{\mathsf\Upsilon_{#1,#2}^{#3}}}
\newcommand{\Dth}[3]{\mathop{\mathsf\Delta_{#1,#2}^{#3}}}
%-------------------------------------------------------------------------------
\begin{defi}\label{DefThrDer}
Consider $n>0$, distinct atoms $a_1$, \dots, $a_n$, and let $p=\lfloor n/2\rfloor$ and $q=n-p$.
\begin{itemize}
%---------------------------------------
%---------------------------------------
\item
For $n>1$ and $1\le l\le n$, we define the derivations $\Uth kln\avec1n$ and $\Dth kln\avec1n$ as follows:
\[
\Uth kln\avec1n=\begin{cases}
\vlinf{\gwu}
      {}
      {\fff}
      {\vls({\vlnos(\th pp\avec1p)}\{a_l/\fff\}.\th{k-p}q\avec{p+1}n)}
             &\text{if $p\le k\le n$ and $l\le p$}\\
\noalign{\medskip}
\vlinf{\gwu}
      {}
      {\fff}
      {\vls(\th{k-q}p\avec1p.{\vlnos(\th qq\avec{p+1}n)}\{a_l/\fff\})}
             &\text{if $q\le k\le n$ and $p<l$}\\
\noalign{\medskip}
\fff         &\text{otherwise}
              \end{cases}
\]
and
\[
\Dth kln\avec1n=\begin{cases}
\vlinf{\gwd}
      {}
      {\th kq\avec{p+1}n}
      {\fff}
             &\text{if $0<k\le q$ and $l\le p$}\\
\noalign{\medskip}
\vlinf{\gwd}
      {}
      {\th kp\avec1p}
      {\fff}
             &\text{if $0<k\le p$ and $p<l$}\\
\noalign{\medskip}
\fff         &\text{otherwise}
              \end{cases}\quad.
\]
%---------------------------------------
%---------------------------------------
\item
For $k\ge0$ and $1\le l\le n$, we define the derivations $\vlsmash{\Gth kln\avec1n}$, recursively on $n$, as follows:
\begin{itemize}
%---------------------------------------
\item $\Gth 011(a_1)=\ttt$;
%---------------------------------------
\item for $k>0$, $\Gth k11(a_1)=\fff$;
%---------------------------------------
\item for $k>n$, $\Gth kln\avec1n=\fff$;
%---------------------------------------
\item for $n>1$ and $k\le n$, let
\[
\Gth kln\avec1n=\begin{cases}
%---------------------------------------
\vls[
\bigvee_{\begin{subarray}{l}i+j=k      \\ 
                            0\le i\lt p\\ 
                            0\le j\le q
         \end{subarray}}(
\Gth ilp\avec1p.
\th jq\avec{p+1}n).
\Uth kln\avec1n.\Dth{k+1}ln\avec1n]
&\text{if $l\le p$}\\
\noalign{\medskip}
%---------------------------------------
\vls[
\bigvee_{\begin{subarray}{l}i+j=k      \\
                            0\le i\le p\\ 
                            0\le j\lt q
         \end{subarray}}(
\th ip\avec1p.
\Gth j{l-p}q\avec{p+1}n).
\Uth kln\avec1n.\Dth{k+1}ln\avec1n]
&\text{if $p<l$}
\end{cases}
\quad.
\]
\end{itemize}
\end{itemize}
\end{defi}

%-------------------------------------------------------------------------------
\begin{exa}
See, in Figure~\ref{FigPThEx}, some example of derivations $\vlsmash{\Gth kln\avec1n}$. Note that, for clarity, we removed all instances of the trivial derivations $\Uth112\avec12=\Uth122\avec12=\Uth113\avec13=\vldownsmash{\vlinf\gwu{}\fff\fff}$. We can do so because these derivation instances appear as disjuncts.
\end{exa}

%-------------------------------------------------------------------------------
\begin{thm}\label{TheoThrDer}
For any $n>0$, $k\ge0$ and\/ $1\le l\le n$, the derivation\/ $\vlsmash{\Gth kln\avec1n}$ has shape
\[
\vlder{}{\{\awd,\awu\}}{(\th{k+1}n\avec1n)\{a_l/\ttt\}}
                       {(\th kn\avec1n)\{a_l/\fff\}}
\quad,
\]
and\/ $\size{\Gth kln\avec1n}$ is $n^{\Ord{\log n}}$.
\end{thm}

%-------------------------------------------------------------------------------
\begin{proof}
The shape of $\Gth kln\avec1n$ can be verified by inspecting Definition~\ref{DefThrDer}. For example, this is the case when $n>1$ and $l\le p\le k<q$, where $p=\lfloor n/2\rfloor$ and $q=n-p$:
\vlstore{\noalign{\medskip}
\vls[
\textstyle\bigvee_{\begin{subarray}{l}i+j=k      \\
                                      0\le i\lt p\\
                                      0\le j\le q
                   \end{subarray}}(
\vlder{\Gth ilp\avec1p}
      {}
      {(\th{i+1}p\avec1p)\{a_l/\ttt\}}
      {(\th ip\avec1p)\{a_l/\fff\}}
.
\th jq\avec{p+1}n)
.
\vlinf{\gwu}
      {}
      {\fff}
      {\vls({\vlnos(\th pp\avec1p)}\{a_l/\fff\}.\th{k-p}q\avec{p+1}n)}
.
\vlinf{\gwd}
      {}
      {\th{k+1}q\avec{p+1}n}
      {\fff}
]}
\begin{multline*}
\vlder{\Gth kln\avec1n}
      {}
      {(\th{k+1}n\avec1n)\{a_l/\ttt\}}
      {(\th kn\avec1n)\{a_l/\fff\}}
={}\\
\vlread
\quad.
\end{multline*}
(Remember that
\[
\th kn\avec1n\equiv\bigvee_{\begin{subarray}{l}
                            i+j=k\\ 
                            0\le i\le p\\ 
                            0\le j\le q
                            \end{subarray}}
                   \vlsbr(\th ip\avec1p.\th jq\avec{p+1}n)
\]
and $\th0p\avec1p\equiv\ttt$.) General (co)weak\-en\-ing rule instances can be replaced by atomic ones because of Proposition~\ref{PropGenAtPol}. The size bound on $\Gth kln\avec1n$ follows from Proposition~\ref{PropGenAtPol} and Theorem~\ref{TheoQuas}.
\end{proof}

%===============================================================================
\section{Normalisation Step 2: Cut-Free Form}\label{SectPreNorm}

In this section we define the cut-free form of proofs, based on proofs in simple form. Proofs in cut-free form have no cut instances, but can have coweakening ones, which prevent these proofs from being analytic (in the sense that atoms appear in premisses only if they do so in conclusions). Theorem~\ref{ThPreNorm}, the main result of the section, shows how to obtain a cut-free proof from any proof. Most of the ingenuity of quasipolynomially normalising an $\SKS$ proof into one in analytic $\SKS$ resides in going from a simple form to a cut-free one. Removing coweakening instances from a cut-free form is easy; we dedicate Section~\ref{SectNorm} to this.

Before defining the cut-free form, we need to establish the following fact.

%-------------------------------------------------------------------------------
\begin{prop}\label{PropAuxNorm}
For any formula $A$ and atom $a$, there exist derivations whose size is cubic in\/ $\size A$ and that have shape
\[
\vlder{}{\{\awd,\acd,\swi\}}{\vls[a.A\{a/\fff\}]}A
\qquad\text{and}\qquad
\vlder{}{\{\awu,\acu,\swi\}}A{\vls(a.A\{a/\ttt\})}
\quad.
\]
\end{prop}

%-------------------------------------------------------------------------------
\begin{proof}
If there are no occurrences of $a$ in $A$, the desired derivations are
\[
\vlderivation            {
\vlin{}{}{\vls[a   .A]} {
\vlin= {}{\vls[\fff.A]}{
\vlhy    A             }}}
\qquad\text{and}\qquad
\vlderivation            {
\vlin= {}A              {
\vlin{}{}{\vls(\ttt.A)}{
\vlhy    {\vls(a   .A)}}}}
\quad.
\]
If there are $h>0$ occurrences of $a$ in $A$, obtain, by repeatedly applying Proposition~\ref{PropSwitch}, the following derivations:
\[
\vlderd{}
       {\{\swi\}}
       {\vlsbr[\vlinf{(h-1)\cdot\acd}
                     {}
                     a
                     {\vls[a.\vldots.a]}.A\{a/\fff\}]}
       A
\qquad\text{and}\qquad
\vlderd{}
       {\{\swi\}}
       A
       {\vlsbr(\vlinf{(h-1)\cdot\acu}
                     {}
                     {\vls(a.\vldots.a)}
                     a.A\{a/\fff\})}
\quad.
\]
If $\size A=n$, the size of the desired derivations is $\Ord{n^3}$ because we have to apply Proposition~\ref{PropSwitch} at most $\Ord n$ times.
\end{proof}

%-------------------------------------------------------------------------------
\begin{figure}
\[
\vcenter{\hbox{\includegraphics{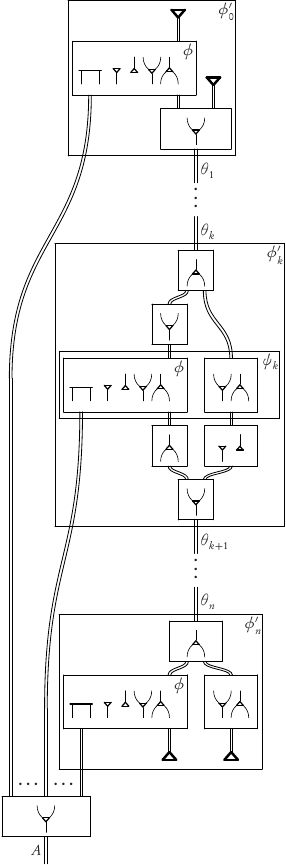}}}
\]
\caption{Atomic flow of a proof in cut-free form.}
\label{FigNormFlow}
\end{figure}

%-------------------------------------------------------------------------------
\begin{defi}\label{DefNorm}
For $n>0$, let $\Pi$ be a proof in simple form over $\avec1n$, such that it and its atomic flow have shape
\[
\hbox{\phantom{$\vls[A.{}]$}}
\vlderd\Psi
       {}
       {\vls[\llap{$\vls[A.{}]$}
             \vlinf{}{}\fff{\vls(a_1.\bar a_1^{\phi_1})}.\cdots.
             \vlinf{}{}\fff{\vls(a_n.\bar a_n^{\phi_n})}]}
       {\vls(\vlinf{}{}{\vls[a_1.\bar a_1^{\phi_1}]}\ttt.\cdots.
             \vlinf{}{}{\vls[a_n.\bar a_n^{\phi_n}]}\ttt)}
\qquad\text{and}\qquad
\vcenter{\hbox{\includegraphics{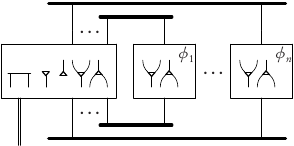}}}
\quad,
\]
for some derivation $\Psi$. For $0\le i\le n+1$, let $\theta_i\equiv\th in\avec1n$. For $0\le k\le n$, we define the derivations
\[
\vlder{\Phi_k}{\SKS\setminus\{\aiu\}}{\vls[A.\theta_{k+1}]}{\theta_k}
\]
as
\[ %%%%% The \dimen's must be adjusted if fonts and layout parameters are changed
\dimen0=3940000sp
\kern\dimen0
\vlderd{\Psi_k}
       {\SKS\setminus\{\aiu\}}
       {\kern-\dimen0\vlsbr[A\vlx.
        \vlinf{\scriptstyle(n-1)\cdot\gcd}
              {}
              {\theta_{k+1}}
              {\vls[\vlderd{}
                           {\{\awu,\acu,\swi\}}
                           {\theta_{k+1}}
                           {\vlsbr(a_1
                                  \vlx.\vlx
                                  \vlder{\Gth k1n\avec1n}
                                        {\{\awd,\awu\}}
                                        {\theta_{k+1}\{a_1/\ttt\}}
                                        {\theta_k    \{a_1/\fff\}})}
                   \vlx.\vlx\vldots\vlx.\vlx
                    \vlderd{}
                           {\{\awu,\acu,\swi\}}
                           {\theta_{k+1}}
                           {\vlsbr(a_n
                                  \vlx.\vlx
                                  \vlder{\Gth knn\avec1n}
                                        {\{\awd,\awu\}}
                                        {\theta_{k+1}\{a_n/\ttt\}}
                                        {\theta_k    \{a_n/\fff\}})}
                   ]}]                                              }  
       {\vlinf{\llap{$\scriptstyle(n-1)\cdot\gcu$}}
              {}
              {\vlsbr(\vlder{}
                            {\{\awd,\acd,\swi\}}
                            {\vls[a_1.\theta_k\{a_1/\fff\}]}
                            {\theta_k                      }
                     \vlx.\vlx\vldots\vlx.\vlx
                      \vlder{}
                            {\{\awd,\acd,\swi\}}
                            {\vls[a_n.\theta_k\{a_n/\fff\}]}
                            {\theta_k                      })}
              {\theta_k                                    }  }
\quad,
\]
where $\Psi_k=\Psi\{\bar a_1^{\phi_1}/\theta_k\{a_1/\fff\},\dots,\bar a_n^{\phi_n}/\theta_k\{a_n/\fff\}\}$ and where we use Proposition~\ref{PropAuxNorm}. We define the \emph{cut-free form of\/ $\Pi$} as the following proof in $\SKS\setminus\{\aiu\}$:
\[ %%%%% The \dimen's must be adjusted if fonts and layout parameters are changed
\dimen0=2350000sp
\dimen1=7570000sp
\dimen2=9980000sp
\vlinf{n\cdot\gcd}
      {\quad.}
      A
      {\kern\dimen2\vlderd{\Phi_0}
             {}
             {\kern-\dimen2
              \vlsbr[A
                    \vlx.\vlx
                    \kern\dimen1
                    \vlderd{\Phi_1}
                           {}
                           {\kern-\dimen1
                            \vlsbr[A
                                  \vlx.\vlx
                                  \cdots\vlx.\vlx\kern\dimen0
                                  \begin{tabular}{@{}c@{}}
                                  $\theta_2$\\
                                  $\vdots$\\
                                  $\kern-\dimen0
                                   \vlsbr[A
                                         \vlx.\vlx
                                         \vlder{\Phi_n}
                                               {}
                                               {\vlsbr[A
                                                      .\theta_{n+1}]}
                                               {\theta_n}]$
                                  \end{tabular}]}
                           {\theta_1}]}
            {\theta_0}}
\]
(We recall that $\theta_0\equiv\ttt$ and $\theta_{n+1}\equiv\fff$.)
\end{defi}

%-------------------------------------------------------------------------------
\begin{thm}\label{ThPreNorm}
Given any proof\/ $\Pi$ of $A$ in\/ $\SKS$, we can construct a proof of $A$ in\/ $\SKS\setminus\{\aiu\}$ in time quasipolynomial in the size of\/ $\Pi$.
\end{thm}

%-------------------------------------------------------------------------------
\begin{proof}
By Theorem~\ref{ThSimpleForm} we can construct from $\Pi$, in polynomial time, a proof $\Pi'$ of $A$ in simple form. We can then proceed with the construction of Definition~\ref{DefNorm}, to which we refer here. For $0\le k\le n$, constructing $\Phi_k$ requires quasipolynomial time because of Propositions~\ref{PropGenAtPol}, \ref{PropSubst} and \ref{PropAuxNorm} and Theorems~\ref{TheoQuas} and \ref{TheoThrDer}, and because obtaining $\Psi_k$ from $\Psi$ requires quasipolynomial time. Constructing the cut-free from of $\Pi'$ from $\Phi_0$, \dots, $\Phi_n$ is done in polynomial time.
\end{proof}

%-------------------------------------------------------------------------------
\begin{rem}
In Figure~\ref{FigNormFlow}, we show the atomic flow of the cut-free form obtained from a proof $\Pi$ in simple form. We refer to Definition~\ref{DefNorm}. Let the following be the flow of the simple core $\Psi$ of $\Pi$:
\[
\vcenter{\hbox{\includegraphics{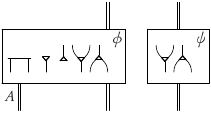}}}\quad,
\]
where $\psi$ is the union of flows $\phi_1$, \dots, $\phi_n$, and where we denote by $A$ the edges corresponding to the atom occurrences appearing in the conclusion $A$ of $\Pi$. We then have that, for $0<k<n$, the flow of $\Phi_k$ is $\phi'_k$, as in Figure~\ref{FigNormFlow}, where $\psi_k$ is the flow of the derivation $\Psi_k$. The flows of $\Phi_0$ and $\Phi_n$ are, respectively, $\phi'_0$ and $\phi'_n$.
\end{rem}

%===============================================================================
\section{Normalisation Step 3: Analytic Form}\label{SectNorm}

Of special importance in this paper is the following proof system:

\newcommand{\aSKS}{\mathsf{aSKS}}
%-------------------------------------------------------------------------------
\begin{defi}
\emph{Analytic\/ $\SKS$} is the system $\aSKS=\SKS\setminus\{\aiu,\awu\}$.
\end{defi}

For example, the system $\{\swi,\med,=,\acd\}$ polynomially simulates the system $\{\swi,=,\gcd\}$, and $\aSKS=\{\swi,\med,=,\aid,\awd,\acd,\acu\}$ polynomially simulates $\{\swi,=,\gid,\gwd,\gcd,\gcu\}$ (where $\gid$ is the nonatomic identity).
In this section, we show that we can get proofs in analytic $\SKS$, \emph{i.e.}, system $\aSKS$, in quasipolynomial time from proofs in $\SKS$.

Transforming a proof in cut-free form into an analytic one requires eliminating coweakening rule instances. This can be done by transformations that are the dual of those over weakening instances, employed in Step~\eqref{ItemWeak} of the proof of Theorem~\ref{ThSimpleForm}.

%-------------------------------------------------------------------------------
\begin{thm}[Jeřábek \cite{Jera::On-the-C:kx}]\label{ThNormAn}
Given any proof\/ $\Pi$ of $A$ in\/ $\SKS$, we can construct a proof of $A$ in\/ $\aSKS$ in time quasipolynomial in the size of\/ $\Pi$.
\end{thm}

%-------------------------------------------------------------------------------
\begin{proof}
By Theorem~\ref{ThPreNorm}, we can obtain, from $\Pi$, a cut-free proof $\Pi'$ of the same formula, in quasipolynomial time in the size of $\Pi$. We associate $\Pi'$ with its atomic flow $\phi$, so that we have a way to identify the atom occurrences in $\Pi'$ associated with each edge of $\phi$, and substitute over them. We repeatedly examine each coweakening instance $\vlinf\awu{}\ttt{a^\epsilon}$ in $\Pi'$, for some edge $\epsilon$ of $\phi$, and we perform one transformation out of the following exhaustive list of cases, for some $\Pi''$, $\Phi$, $\Psi$, $K\vlhole$ and $H\vlhole$:
\begin{enumerate}
%---------------------------------------
\item\label{ItemIDWU}
\[
\vlderivation                                             {
\vldd{\Psi }{}A                                          {
\vldd{\Phi }{}{H{\left\{\vlinf{}
                              {}
                              \ttt
                              {a^\epsilon}\right\}}    }{
\vlpd{\Pi''}{}{K{\left\{\vlinf{}
                              {}
                              {\vls[a^\epsilon.\bar a]}
                              \ttt\right\}}            }}}}
\qquad\text{becomes}\qquad
\vlderivation                                 {
\vlde{\Psi }{}A                              {
\vldd{\Phi\{a^\epsilon/\ttt\}
           }{}{H\vlscn(\ttt)               }{
\vlpd{\Pi''}{}{K{\vlsbr[\ttt.\vlinf{}
                                   {}
                                   {\bar a}
                                   \fff]}  }}}}
\quad;
\]
%---------------------------------------
\item\label{ItemWDWU}
\[
\vlderivation                                         {
\vldd{\Psi }{}A                                      {
\vldd{\Phi }{}{H{\left\{\vlinf{}
                              {}
                              \ttt
                              {a^\epsilon}\right\}}}{
\vlpd{\Pi''}{}{K{\left\{\vlinf{}
                              {}
                              {a^\epsilon}
                              \fff\right\}}        }}}}
\qquad\text{becomes}\qquad
\vlderivation                                                     {
\vlde{\Psi }{}A                                                  {
\vldd{\Phi\{a^\epsilon/\ttt\}
           }{}{H\vlscn(\ttt)                                   }{
\vlpd{\Pi''}{}{K{\left\{\vlinf\swi
                              {}
                              {\vls[(\fff.\ttt).\ttt]}
                              {\vls(\fff.[\ttt.\ttt])}\right\}}}}}}
\quad;
\]
%---------------------------------------
\item\label{ItemCDWU}
\[
\vlderivation                                         {
\vldd{\Psi }{}A                                      {
\vldd{\Phi }{}{H{\left\{\vlinf{}
                              {}
                              \ttt
                              {a^\epsilon}\right\}}}{
\vlpd{\Pi''}{}{K{\left\{\vlinf{}
                              {}
                              {a^\epsilon}
                              {\vls[a.a]}\right\}} }}}}
\qquad\text{becomes}\qquad
\vlderivation                        {
\vlde{\Psi }{}A                     {
\vldd{\Phi\{a^\epsilon/\ttt\}
           }{}{H\vlscn(\ttt)      }{
\vlpd{\Pi''}{}{K{\vlsbr[\vlinf{}
                              {}
                              \ttt
                              a
                       .\vlinf{}
                              {}
                              \ttt
                              a]} }}}}
\quad;
\]
%---------------------------------------
\item\label{ItemCUWU}
\[
\vlderivation                                         {
\vldd{\Psi}{}A                                       {
\vldd{\Phi }{}{H{\left\{\vlinf{}
                              {}
                              \ttt
                              {a^\epsilon}\right\}}}{
\vlpd{\Pi''}{}{K{\left\{\vlinf{}
                              {}
                              {\vls(a^\epsilon.a)}
                              a\right\}}           }}}}
\qquad\text{becomes}\qquad
\vlderivation                  {
\vlde{\Psi}{}A                {
\vlde{\Phi\{a^\epsilon/\ttt\}
           }{}{H\vlscn(\ttt)}{
\vlpr{\Pi''}{}{K\vlscn(a)   }}}}
\quad.
\]
\end{enumerate}
The process terminates in linear time on the size of $\Pi'$ because each transformation eliminates some atom occurrences. The final proof is in $\aSKS$.
\end{proof}

The transformations described in the proof of Theorem~\ref{ThNormAn} are the minimal ones necessary to produce a proof in $\aSKS$. However, it is possible to further reduce the proof so obtained. The transformations in the proof of Theorem~\ref{ThNormAn}, together with the one mentioned in Step~\eqref{ItemWeak} in the proof of Theorem~\ref{ThSimpleForm}, all belong to the class of weakening and coweakening reductions studied in
\cite{GuglGund:07:Normalis:lr}. In the rest of this section, we quickly outline a possible, further transformation of the analytic form produced by those reductions, and refer the reader to
\cite{GuglGund:07:Normalis:lr} for a more thorough explanation.

It is advantageous to describe the weakening and coweakening transformations directly as atomic-flow reduction rules. These are special graph rewriting rules for atomic flows, that are known to correspond to sound derivation transformations, in the following sense. If $\Phi$ is a derivation with flow $\phi$, and $\phi$ can be transformed into $\psi$ by one of the atomic-flow reduction rules, then there exists a derivation $\Psi$ whose flow is $\psi$ and such that it has the same premiss and conclusion as $\Phi$. Moreover, $\Psi$ can be obtained from $\Phi$ by instantiating some atoms and changing some rule instances, in linear time.

\newcommand{\rwdcd}
   {{{\mathsf{aw}}{\downarrow}{\hbox{-}}{\mathsf{ac}}{\downarrow}}}
\newcommand{\rwdiu}
   {{{\mathsf{aw}}{\downarrow}{\hbox{-}}{\mathsf{ai}}{\uparrow  }}}
\newcommand{\rwdwu}
   {{{\mathsf{aw}}{\downarrow}{\hbox{-}}{\mathsf{aw}}{\uparrow  }}}
\newcommand{\rwdcu}
   {{{\mathsf{aw}}{\downarrow}{\hbox{-}}{\mathsf{ac}}{\uparrow  }}}
\newcommand{\rcuwu}
   {{{\mathsf{ac}}{\uparrow  }{\hbox{-}}{\mathsf{aw}}{\uparrow  }}}
\newcommand{\rcdwu}
   {{{\mathsf{ac}}{\downarrow}{\hbox{-}}{\mathsf{aw}}{\uparrow  }}}
\newcommand{\ridwu}
   {{{\mathsf{ai}}{\downarrow}{\hbox{-}}{\mathsf{aw}}{\uparrow  }}}
The weakening and coweakening atomic-flow reduction rules are shown in Figure~\ref{FigWeakRed}. The reduction rule labelled $\rwdiu$ is employed in Step~\eqref{ItemWeak} in the proof of Theorem~\ref{ThSimpleForm}. The reduction rules labelled $\rcuwu$, $\ridwu$, $\rwdwu$ and $\rcdwu$ are employed in the proof of Theorem~\ref{ThNormAn}, respectively as Case~\eqref{ItemCUWU}, \eqref{ItemIDWU}, \eqref{ItemWDWU} and \eqref{ItemCDWU}. If we apply the full set of weakening and coweakening reductions until possible, starting from a proof in cut-free form, we obtain a proof of the same formula and whose flow has shape
\[
\vcenter{\hbox{\includegraphics{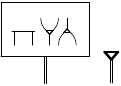}}}\quad.
\]
Note that the graph rewriting system consisting of the reductions in Figure~\ref{FigWeakRed} is confluent.

%-------------------------------------------------------------------------------
\begin{figure}
\[
\begin{array}{@{}c@{}c@{}}
%-------------------
\rwdcd\colon\quad
\vcenter{\hbox{\includegraphics{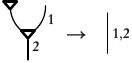}}}
&\qquad\qquad
%-------------------
\rcuwu\colon\quad
\vcenter{\hbox{\includegraphics{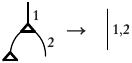}}}
\\
\noalign{\bigskip}
%-------------------
\rwdiu\colon\quad
\vcenter{\hbox{\includegraphics{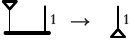}}}
&\qquad\qquad
%-------------------
\ridwu\colon\quad
\vcenter{\hbox{\includegraphics{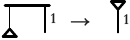}}}
\\
\noalign{\bigskip}
%-------------------
\multispan2{\hfil$
\rwdwu\colon\quad
\vcenter{\hbox{\includegraphics{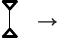}}}
$\hfil}\\
\noalign{\bigskip}
%-------------------
\rwdcu\colon\quad
\vcenter{\hbox{\includegraphics{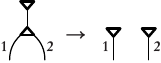}}}
&\qquad\qquad
%-------------------
\rcdwu\colon\quad
\vcenter{\hbox{\includegraphics{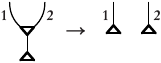}}}
\end{array}
\]
\caption{Weakening and coweakening atomic-flow reductions.}
\label{FigWeakRed}
\end{figure}

%===============================================================================
\section{Final Comments and Future Work}\label{SectFinComm}

\newcommand{  \LK}{\mathsf{  LK}}
\newcommand{ \MLK}{\mathsf{ MLK}}
\newcommand{\MCLK}{\mathsf{MCLK}}

System $\aSKS$ is not a minimal complete system for propositional logic, because the atomic cocontraction rule $\acu$ is admissible (via $\acd$, $\swi$, $\aiu$ and cut elimination). Removing $\acu$ from $\aSKS$ yields system $\KS$. A natural question is whether quasipolynomial-time normalisation holds for $\KS$ as well. We would guess that cocontraction plays an essential role in keeping the complexity low. For example, one can note in Figure~\ref{FigNormFlow} how cocontraction limits the size of the $n$ pieces of derivation below each $\theta_k$. If we had to expand those cocontraction instances into a tree we would have an exponential blow-up. On the other hand, an encouraging result in the opposite direction is contained in \cite{Das:13:The-Pige:fk}, where the author obtains $n^{O(\log\log n)}$-size proofs of the weak pigeonhole principle, using deep-inference techniques to improve the previous bound for monotone proofs.

There is reason to believe that polynomial normalisation is achievable, because it is possible to compute threshold functions with polynomial formulae. However, the hardest problem seems to be obtaining polynomial $\Gammasf$-like (cut-free) derivations with the property of Theorem~\ref{TheoThrDer}. We tend to think that polynomiality ought to be possible, and deep inference might be a helpful language for investigating and achieving it, because of its flexibility in constructing derivations.

The normalisation procedure presented here is peculiar because it achieves its result by using an external scheme, constituted by the threshold formulae and the $\Gammasf$ derivations, which does not depend on the derivation to be normalised. Threshold formulae realise a clever compositional mechanism built on top of cocontractions. It would be interesting to interpret this or a similar mechanism computationally. We do not necessarily expect an interpretation strictly following the Curry-Howard scheme, and there is little evidence that the threshold construction studied here can be applied in the intuitionistic case. On the other hand an essential ingredient of our construction, namely sharing by cocontractions, already allowed us to reach beyond the limitations of Gentzen-style proof theory: in recent work fully lazy sharing has been achieved in a $\lambda$-calculus within intuitionistic deep inference \cite{GundHeijPari:13:Atomic-L:fk,GundHeijPari:13:A-Proof-:kx}. For the authors of this paper and for some of their colleagues this is an active research area.

It is possible to extend the mechanism investigated here to the more general notion of normalisation that we called \emph{streamlining} in \cite{GuglGund:07:Normalis:lr}; this has been done in \cite{Gund:09:A-Genera:kx}. Streamlining is a top-down symmetric notion, that does full justice to the additional symmetry of deep inference, compared to Gentzen formalisms. Streamlined derivations entail analytic proofs as a special case.

The results of this paper are, as mentioned previously, closely related to results about the monotone sequent calculus $\MLK$, through the translation between $\SKS$ and $\LK$ given in \cite{Brun:06:Deep-Inf:qy}. Atserias, Galesi and Pudlák show in \cite{AtseGalePudl:02:Monotone:yu} that $\MLK$ can quasipolynomially simulate $\LK$ over monotone sequents, and as shown by Jeřábek in \cite{Jera::On-the-C:kx}, this implies Theorem~\ref{ThPreNorm}.

Furthermore, in \cite{Jera:12:Proofs-w:fk}, Jeřábek considers a conservative extension of $\MLK$, called $\MCLK$, and shows how $\MCLK$ can quasipolynomially simulate $\LK$ over arbitrary formulae. $\MCLK$ is defined by restricting $\LK$ to only allow cuts on monotone formulae. Like $\MCLK$, streamlined derivations are also a conservative extension of $\MLK$ and the two notions are very similar. Exploring this will be the subject of future work.

Finally, we are interested in the normalisation theory of modal logics in deep inference, and so we are naturally led to consider the methods presented in this paper to that purpose as well.

%===============================================================================

\vspace{-30 pt}
\end{document}